\documentclass[a4paper,11pt]{article}
\pdfoutput=1
\usepackage{jheppub}
\usepackage{tikz}

\usepackage{enumitem}
\usepackage[T1]{fontenc} 
\usepackage{floatrow}
\usepackage{appendix}
\usepackage{tabularx}
\usepackage[normalem]{ulem}

\allowdisplaybreaks

\usepackage{epsf}
\usepackage{amsmath}
\usepackage{amsfonts}
\usepackage{amssymb}
\usepackage{psfrag,epsfig,graphicx,graphics}

\newcommand\numberthis[1][]{%
    \refstepcounter{equation}%
    \ifx#1\empty\else\label{eq:#1}\fi%
    \tag{\theequation}%
}

\usepackage{xargs} 
\usepackage[colorinlistoftodos,prependcaption,textsize=tiny]{todonotes}
\newcommandx{\AP}[2][1=]{\todo[linecolor=blue,backgroundcolor=blue!25,bordercolor=blue,#1]{#2}}
\newcommandx{\LDR}[2][1=]{\todo[linecolor=black,backgroundcolor=black!25,bordercolor=black,#1]{#2}}
\newcommandx{\GG}[2][1=]{\todo[linecolor=orange,backgroundcolor=orange!25,bordercolor=orange,#1]{#2}}

\newcommandx{\FGC}[2][1=]{\todo[linecolor=red,backgroundcolor=red!25,bordercolor=red,#1]{#2}}

\newcommandx{\MF}[2][1=]{\todo[linecolor=brown,backgroundcolor=brown!25,bordercolor=brown,#1]{#2}}

\providecommand{\U}[1]{\protect\rule{.1in}{.1in}}

\def\slashchar#1{\setbox0=\hbox{$#1$}
   \dimen0=\wd0
   \setbox1=\hbox{/} \dimen1=\wd1
   \ifdim\dimen0>\dimen1
      \rlap{\hbox to \dimen0{\hfil/\hfil}}
      #1
   \else
      \rlap{\hbox to \dimen1{\hfil$#1$\hfil}}
      /
   \fi}

\def\nn{\nonumber}

\def\bei{\begin{itemize}}
\def\ei{\end{itemize}}

\def\beeq{\begin{eqnarray}} 
\def\beqa{\begin{eqnarray}}
\def\bea{\begin{eqnarray}}

\def\eea{\end{eqnarray}}
\def\eqa{\end{eqnarray}}
\def\eeeq{\end{eqnarray}}

\def\eqar{\end{array}}
\def\beqar{\begin{array}}

\def\beas{\begin{eqnarray*}}
\def\beqas{\begin{eqnarray*}}

\def\eqas{\end{eqnarray*}}
\def\eeas{\end{eqnarray*}}

\def\beq{\begin{equation}} 
\def\be{\begin{equation}}

\def\ee{\end{equation}}
\def\eq{\end{equation}}
\def\eeq{\end{equation}}

\def\beqd{\begin{displaymath}}
\def\eeqd{\end{displaymath}}
\def\eqd{\end{displaymath}}

\def\beeq{\begin{eqnarray}} \def\eeeq{\end{eqnarray}}

\newcommand{\fin}{\end{document}}

\title{\boldmath The next-to-leading order Higgs impact factor at physical top mass: The real corrections}

\author[a]{Francesco Giovanni Celiberto,}
\author[b,c]{Luigi Delle Rose,}
\author[d,1]{Michael Fucilla,\note{Corresponding author.}}
\author[b,c]{Gabriele Gatto,}
\author[b,c]{Alessandro Papa}

\affiliation[a]{Universidad de Alcalá (UAH), Departamento de Física y Matemáticas, Campus Universitario, Alcalá de Henares, E-28805, Madrid, Spain}

\affiliation[b]{Dipartimento di Fisica, Università della Calabria, Arcavacata di Rende, I-87036, Cosenza, Italy}

\affiliation[c]{
INFN, Gruppo Collegato di Cosenza, Arcavacata di Rende, I-87036, Cosenza, Italy}

\affiliation[d]{Université Paris-Saclay, CNRS/IN2P3, IJCLab, 91405, Orsay, France}

\emailAdd{francesco.celiberto@uah.es}
\emailAdd{luigi.dellerose@unical.it}
\emailAdd{michael.fucilla@ijclab.in2p3.fr}
\emailAdd{gabriele.gatto@unical.it}
\emailAdd{alessandro.papa@fis.unical.it}

\abstract{We compute the real corrections to the impact factor for the production of a forward Higgs boson, retaining full top-mass dependence. We demonstrate that the rapidity divergence is the one predicted by the BFKL factorization and perform the explicit subtraction in the BFKL scheme. We show that the IR-structure of the impact factor is the expected one and that, in the infinite-top-mass approximation, the previously known result is recovered. We also verify that the impact factor vanishes when the transverse momenta of the $t$-channel Reggeon goes to zero, in agreement with its gauge-invariant definition, exploiting the $m_t \rightarrow \infty$ expansion up to the next-to-next-to-leading order.}

\begin{document} 
\maketitle
\flushbottom

\section{Introduction}
\label{sec:intro}

The long-awaited discovery of the Higgs boson at the LHC opened a breach in an unexplored sector of the Standard Model, thus offering a new stage for the quest of New Physics, {\it via} the comparison of novel experimental analyses with more and more precise predictions.
The established framework for theoretical investigations in the Higgs sector in hadron-hadron collisions is {\it collinear factorization} (see Ref.~\cite{Collins:1989gx} for a review), which is based on the convolution of universal, non-perturbative parton distribution functions (PDFs) with perturbative, process-dependent coefficient functions, whose precise determination relies mostly upon the inclusion of QCD radiative corrections beyond the leading order (LO). The current status is next-to-next-to-next-LO (or N$^3$LO) in the infinite-top-mass limit~\cite{Anastasiou:2015vya,Mistlberger:2018etf}, where the gluon-Higgs interaction is described by a dimension-5 effective operator. At physical top mass the current state of the art is next-to-next-to-leading order (NNLO)~\cite{Czakon:2021yub}. \\

There are however some corners in the final-state phase space where collinear factorization alone is not able to reach the required precision level, because large logarithms of some kinematic variable appear, which compensate the smallness of the strong coupling and must therefore be resummed to all perturbative orders. This is the case of the region of small transverse momenta, where Higgs distributions can be consistently described if transverse-momentum (TM) resummation is properly considered~\cite{Catani:2000vq,Bozzi:2005wk,Catani:2011kr,Catani:2013tia,Monni:2019yyr}. Another interesting domain is the one where Higgs production originates from partons carrying large fractions of the proton momentum~\cite{Bonvini:2016wki,Bonvini:2018ixe}, the so-called large-$x$ sector, with $x$ being the Bjorken variable. Here, large (Sudakov) double logarithms arising from soft and collinear gluon emissions at large $x$ must be all-order resummed, the related theoretical apparatus being provided by the {\it threshold} resummation~\cite{Sterman:1986aj,Catani:1996yz,Ahrens:2009cxz,deFlorian:2012yg,Ball:2013bra,Bonvini:2014joa,Muselli:2017bad,Forte:2021wxe}. Taking advantage of the fact that the small-$x$ resummation and the large-$x$ one respectively control the two opposite tails of the cross section projected in the Mellin space, a pioneering double and joint resummation of both high-energy and threshold logarithms was achieved in Ref.~\cite{Bonvini:2018ixe} for inclusive-Higgs hadroproduction rates. \\

The kinematic regime relevant for the present work is the small-$x$ one or, equivalently, the regime where the center-of-mass-energy $s$ is large with respect to the hard scale of the process ({\it i.e.} the transverse Higgs mass), which is in its turn large with respect to the QCD mass scale, $\Lambda_{\rm QCD}$. This is the so-called {\it semi-hard} regime~\cite{Gribov:1984tu} (for a review of novel applications, see Refs.~\cite{Celiberto:2022dyf,Hentschinski:2022xnd,Celiberto:2022rfj,Celiberto:2017ius,Mohammed:2022gbk} and references therein), where the large logarithms to be all-order resummed are the linear ones in the energy. This is the regime where the Higgs boson is inclusively produced at forward rapidities, possibly in association with a backward emitted object, a jet or a hadron. The theoretical framework for this resummation is provided by the Balitsky-Fadin-Kuraev-Lipatov (BFKL)~\cite{Fadin:1975cb,Kuraev:1976ge,Kuraev:1977fs,Balitsky:1978ic} approach, which was developed in the leading-logarithmic approximation (LLA), {\it i.e.} resummation of all terms proportional to $[\alpha_s\ln(s)]^n$, and in the next-to-LLA (NLLA), {\it i.e.} resummation of all terms of the form $\alpha_s[\alpha_s\ln(s)]^n$.
%At the basis of the BFKL framework is the property of gluon Reggeization %in QCD~\cite{Grisaru:1973vw,Grisaru:1973wbb,Lipatov:1976zz}, which means %that there is a Reggeon with gluon quantum numbers, negative signature %and trajectory $j(t) = 1 + \omega(t)$ passing through 1 at $t = 0$, %which gives the leading contribution  to amplitudes with gluon quantum %numbers in the $t$-channel, in each order of perturbation theory.
%This remarkable property appeared first in direct calculations at fixed %order~\cite{Lipatov:1976zz,Kuraev:1976ge}. Then, it was proved, both in %the LLA~\cite{Balitsky:1979ap} and in the NLLA (see~\cite{Fadin:2015ym} %and references therein) using bootstrap relations~\cite{Fadin:2006pr} %following from the requirement of compatibility of the pole Regge form %with the $s$-channel unitarity. 
In the BFKL approach, (differential) cross sections take the peculiar form of a convolution, in transverse-momenta space, of two process-dependent impact factors, describing the transition of each colliding particle to a definite state in its fragmentation region, and a universal, process-independent Green's function, which encodes the resummation of energy logarithms. The BFKL Green's function is determined by an integral equation, whose kernel is known up to next-to-leading order (NLO), both for forward scattering ({\it i.e.} for $t = 0$ and color singlet in the $t$-channel)~\cite{Fadin:1998py,Ciafaloni:1998gs} and for any fixed, not growing with $s$, momentum transfer $t$ and any possible two-gluon colored exchange in the $t$-channel~\cite{Fadin:1998jv,Fadin:2000kx,Fadin:2000hu,Fadin:2004zq,Fadin:2005zj}.
The calculation of pieces of the next-to-NLO kernel has been the object of recent investigations in $\mathcal{N}=4$ SYM~\cite{Byrne:2022wzk}, in pure-gauge QCD~\cite{DelDuca:2021vjq} and in full QCD~\cite{Caola:2021izf,Falcioni:2021dgr,Fadin:2023roz,Abreu:2024mpk}. As for impact factors, only a few of them are known with NLO accuracy, where their calculation is usually quite arduous: (i) quark and gluon impact factors~\cite{Fadin:1999de,Fadin:1999df,Ciafaloni:1998kx,Ciafaloni:1998hu,Ciafaloni:2000sq}, which are closely related to the (ii) forward-jet~\cite{Bartels:2001ge,Bartels:2002yj,Caporale:2011cc,Ivanov:2012ms,Colferai:2015zfa} and (iii) forward light-hadron~\cite{Ivanov:2012iv} impact factors, (iv) the impact factor for the light vector-meson electroproduction~\cite{Ivanov:2004pp}, (v) the ($\gamma^* \to \gamma^*$) impact factor~\cite{Bartels:2000gt,Bartels:2001mv,Bartels:2002uz,Bartels:2003zi,Bartels:2004bi,Fadin:2001ap,Balitsky:2012bs}. Recently, the one-loop corrections to the (vi) photon-initiated and (vii) gluon-initiated S-wave heavy-quarkonium impact factors have been computed~\cite{Nefedov:2023uen,Nefedov:2024swu} within the Lipatov effective-field theory (EFT) framework~\cite{Lipatov:1995pn}.
Relevant for Higgs physics is (viii) the impact factor for the forward-Higgs production from an incoming proton, which was calculated so far in the infinite-top-mass limit~\cite{Nefedov:2019mrg,Hentschinski:2020tbi,Celiberto:2022fgx,Fucilla:2022whr,Fucilla:2023pma,Celiberto:2023dkr,Celiberto:2023rqp}. A lot of predictions have been built, by combining NLO impact factors with the NLLA Green's function, or with only a partial inclusion of NLLA effects, by convoluting the NLLA Green's with one or both the impact factors taken at the leading order (LO), up to the NLO corrections dictated by renormalization group invariance (see, {\it e.g.}, \cite{Celiberto:2020wpk,Celiberto:2022keu,Celiberto:2024omj} and references therein).
So far, BFKL predictions concerning forward Higgs productions have been
obtained in partial NLLA and were concerned with the inclusive production of a Higgs plus a rapidity-separated jet~\cite{Celiberto:2020tmb,Celiberto:2023rtu} or charmed hadron~\cite{Celiberto:2022zdg}; more recently, Higgs-plus-jet predictions were complemented by matching procedures with fixed-order calculations~\cite{Andersen:2022zte,Andersen:2023kuj,Celiberto:2023uuk,Celiberto:2023eba,Celiberto:2023nym}). The extension of these predictions to the full NLLA case is now possible, at least in the infinite-top-mass approximation. In the same approximation, full BFKL NLLA predictions are within reach for the inclusive single forward Higgs production, to be designed by convoluting the NLO Higgs impact factor with a suitably defined {\it unintegrated gluon distribution}~(see for instance~\cite{Hentschinski:2012kr,Hentschinski:2013id,Besse:2013muy,Chachamis:2015ona,Bolognino:2018rhb,Bolognino:2019pba,Bolognino:2021niq}). \\

Relaxing the infinite-top-mass approximation and calculating the NLO Higgs impact factor at the physical top mass is a mandatory step forward, for, at least, two reasons. The first of them is quite obvious and is the urge for precision: top-mass effects are expectedly non-negligible when the Higgs transverse momentum approaches the top mass (see for instance~\cite{Maltoni:2018dar}). Embodying heavy-quark finite mass corrections in the high-energy resummation can represent a valuable tool for explorations at larger center-of-mass energies, such as the nominal ones of the Future Circular Collider FCC~\cite{FCC:2018byv,FCC:2018evy,FCC:2018vvp,FCC:2018bvk}. Here, not only small-$x$ effects are heightened~\cite{Bonvini:2018ixe}, but even bottom-mass effects might become relevant~\cite{Forte:2015hba,Forte:2016sja,Bonciani:2022jmb}.
The second reason is less obvious and is related to some formal subtleties
which arose in the calculation of the NLO Higgs impact factor in the
infinite-top-mass limit and were addressed in a recent work~\cite{Fucilla:2024cpf}. It was found that the usual eikonal approximation (Gribov prescription), which is a common tool in the calculation of 
high-energy amplitudes and BFKL impact factors, actually breaks down
in the case of the NLO Higgs impact factor in infinite-top-mass limit
and some non-eikonal terms must also be included. Moreover, another common technique in this context, the method of expansion in rapidity regions, produces unexpected terms which break the standard rapidity partitioning. Both issues can be traced back to the fact that, in the infinite-top mass limit, the gluon-Higgs coupling is described by a non-renormalizable dimension-5 operator. Interestingly, in 
Ref.~\cite{Fucilla:2024cpf} it was found that non-eikonal terms and 
terms breaking the rapidity partitioning cancel each other exactly, thus restoring the expectations based on the standard Regge form of high-energy amplitudes. It would be interesting to check if the same formal subtleties appear when calculating the NLO Higgs impact factor for finite top mass. \\

This paper is a first step towards fully addressing these issues, presenting the calculation of the {\it real corrections} of the NLO Higgs impact factor for finite top mass. These NLO corrections are those related to the emission of an extra parton in the same fragmentation region of the projectile where the Higgs is produced. The scattering amplitudes necessary for this calculation were computed for the first time in Refs.~\cite{DelDuca:2001fn,DelDuca:2001eu} (see also Ref.~\cite{Budge:2020oyl}). We use Ref.~\cite{Baur:1989cm,Rozowsky:1997dm,DelDuca:2001fn} to cross-check our results. The next and final step will be the calculation of {\it virtual corrections}, which will be considered in a future publication. The set of two-loop master integrals for the Higgs-plus-jet production has been provided in Refs.~\cite{Bonciani:2016qxi,Bonciani:2019jyb,Frellesvig:2019byn} and recently used to compute the Higgs-plus-jet cross-section at NLO in QCD with full top- and bottom-mass dependence~\cite{Bonciani:2022jmb}. The comparison with the expected Regge form of the large center-of-mass energy limit of the two-loop amplitudes, used in~\cite{Bonciani:2022jmb}, will enable to extract the virtual corrections to the forward Higgs impact factor. \\

The outline of the paper is as follows: in Section~\ref{sec:theory}, we recall the basics of the BFKL approach; in Section~\ref{sec:amplitudes}, we present the calculation of the off-shell amplitudes which will be needed in later sections; in Section~\ref{sec:kinematicsLO}, we give the kinematics and the LO calculation; in Section~\ref{sec:NLOcomputation}, we present the NLO real corrections, separately for the gluon- and the quark-initiated subprocess; in Section~\ref{sec:conclusions}, we draw our conclusions.

\section{Theoretical framework}
\label{sec:theory}

\subsection{Generalities of the BFKL approach}
\label{ssec:BFKL_general}

We briefly recall here the generalities of the BFKL approach, starting our discussion from the fully inclusive parton-parton cross section $A(k_A)+B(k_B) \to {\rm all}$, which, through the optical theorem, can be related to the $s$-channel imaginary part of the 
elastic amplitude $A(k_A)+B(k_B) \to A(k_A)+B(k_B)$ at zero transferred momentum,
\begin{equation}
    \sigma_{AB} = \frac{\Im m_{s} \mathcal{A}}{s}\;, 
\end{equation}
with $s = ({k}_A + {k}_B)^2$.
The BFKL approach is introduced to describe this cross section in the limit $s\to \infty$.

We use for all vectors the Sudakov decomposition
\begin{equation}
p = \beta k_1 + \alpha k_2 + p_{\perp},\ \ \ \ \ \ \ \ p_{\perp}^2 =
- \vec p^{\:2}~,
\label{Sudakov1}
\end{equation}
the vectors $(k_1,\ k_2)$ being the light-cone basis of the initial
particle momenta plane $(k_A,\ k_B)$, so that we can put
\begin{equation}
k_A= k_1 +
\frac{m_A^2}{s}k_2~,\ \ \ \ \ \ \ k_B =k_2 +
\frac{m_B^2}{s} k_1~.
\label{Sudakov2}
\end{equation}
Here $m_A$ and $m_B$ are the masses of the colliding partons $A$ and $B$ (taken equal to zero) and the vector notation is used throughout this paper for the transverse
components of the momenta, since all vectors in the transverse subspace are 
evidently space-like.
\begin{figure}
    \centering
    \includegraphics[scale=0.50]{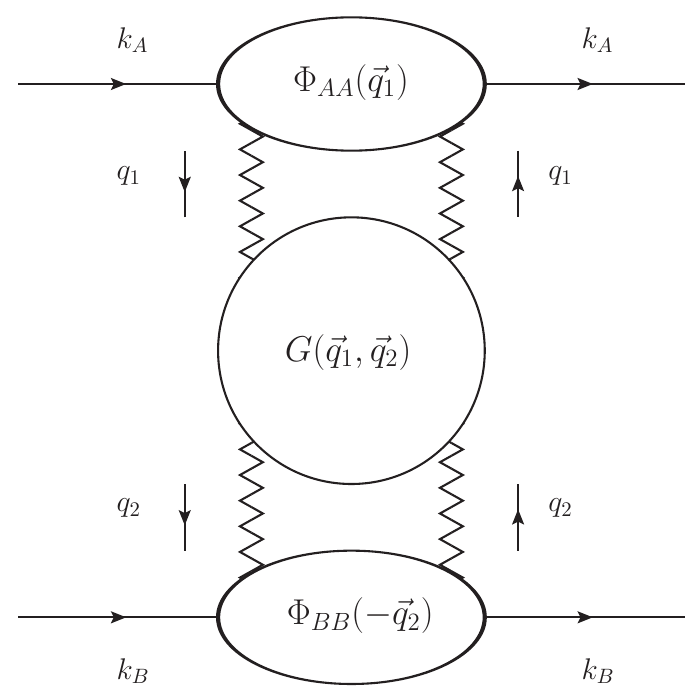}
    \caption{Diagrammatic representation of the elastic scattering amplitude $A + B \rightarrow A + B$.}
\label{GreenCon}
\end{figure}
Within the NLLA, the elastic scattering amplitude for $t=0$ can be written as
\[
\Im m_{s}\left(({\cal A})_{AB}^{AB}\right) = \frac{s}{\left( 2\pi \right)^{D-2}}
\int \frac{d^{D-2}q_1}{\vec{q}_{1}^{\:2}}
\int \frac{d^{D-2}q_2}{\vec{q}_{2}^{\:2}}
\]
\begin{equation}
\times \Phi _{AA}\left( \vec{q}_{1};s_{0}\right)\int_{\delta -i\infty}^{\delta+i\infty}
\frac{d\omega }{2\pi i}\left[ \left( \frac{s}{s_{0}}\right)^{\omega }
G_{\omega }\left( \vec{q}_{1},\vec{q}_{2}\right) 
\right] \Phi _{BB}\left( -\vec{q}_{2};s_{0}\right) \;,
\label{Ar}
\end{equation}
where momenta are defined in Fig.~\ref{GreenCon} and $D$ is the space-time dimension, which
will be taken equal to $4 - 2 \epsilon$ in order to regularize divergences. 
In the above equation $\Phi_{P P}$ are the impact factors and $G_{\omega}$ is the Mellin transform of the Green's function for the Reggeon-Reggeon 
scattering~\cite{Fadin:1998fv}. 
The parameter $s_0$ is an arbitrary energy scale introduced in order to define the partial wave expansion of the scattering amplitudes. The dependence 
on this parameter disappears in the full expressions 
for the amplitudes, within NLLA. The integration in the complex plane $\omega$ is performed along the line $\Re e(\omega)=\delta$ which is supposed to lie to the right of all
singularities in $\omega$ of $G_\omega$.

The Green's function obeys the BFKL equation
\begin{equation}
\omega G_{\omega }\left( \vec{q}_{1},\vec{q}_{2}\right) = \delta^{\left(D-2\right) }\left( \vec{q}_{1}-\vec{q}_{2}\right)
+\int d^{D-2} q_r \ {\cal K}\left( \vec{q}_{1},\vec{q}_r\right) G_{\omega }\left( \vec{q}_r,\vec{q}_{2}\right) \;,
\label{genBFKL}
\end{equation}
where ${\cal K}$ is the NLO kernel in the singlet color representation~\cite{Fadin:1998fv}. The definition of NLO impact factor can be found in Ref.~\cite{Fadin:1998fv} and for $t=0$ reads
$$
\Phi_{AA}(\vec q_1; s_0) = \left( \frac{s_0}
{\vec q_1^{\:2}} \right)^{\omega( - \vec q_1^{\:2})}
\sum_{\{f\}}\int\theta(s_{\Lambda} -
s_{AR})\frac{ds_{AR}}{2\pi}\ d\rho_f \ \Gamma_{\{f\}A}^c
\left( \Gamma_{\{f\}A}^{c^{\prime}} \right)^* 
\langle cc^{\prime} | \hat{\cal P}_0 | 0 \rangle
$$
\begin{equation}
-\frac{1}{2}\int d^{D-2}q_2\ \frac{\vec q_1^{\:2}}{\vec q_2^{\:2}}
\: \Phi_{AA}^{(0)}(\vec q_2)
\: {\cal K}^{(0)}_r (\vec q_2, \vec q_1)\:\ln\left(\frac{s_{\Lambda}^2}
{s_0(\vec q_2 - \vec q_1)^2} \right)~,
\label{ImpactUnpro}
\end{equation}
where $\omega(t)$ is the Reggeized gluon trajectory, which enters this expression at the LO, given by
\begin{equation}
    \omega^{(1)}(t) = \frac{g^2 t}{(2 \pi)^{D-1}} \frac{N}{2} \int \frac{d^{D-2}k_{\perp}}{k_{\perp}^2 (q-k)_{\perp}^{2}} = - \frac{g^2 N \Gamma(1+\epsilon)(\vec{q}^{\; 2})^{-\epsilon}}{(4 \pi)^{2-\epsilon}} \frac{\Gamma^2(-\epsilon)}{\Gamma(-2\epsilon)} \; ,
    \label{ReggeTraj}
\end{equation}
with $t=q^2=-\vec q^{\:2}$ and $N$ the number of colors.
\begin{figure}
\begin{center}
\includegraphics[scale=0.50]{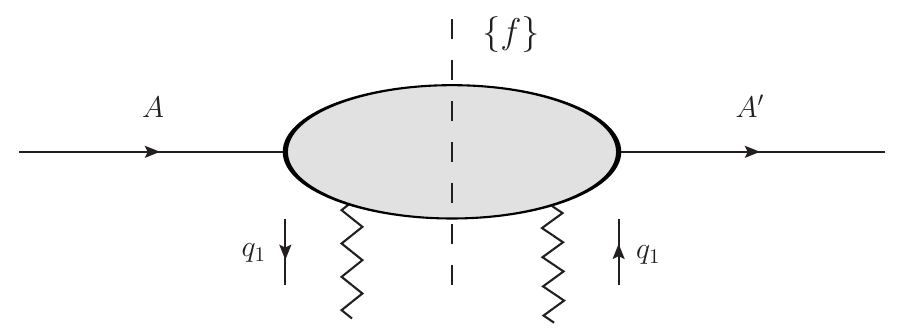}
\end{center}
\caption{Schematic description of an impact factor.}
\label{ImpactFactorRepres}
\end{figure}
$\Gamma_{\{f\}A}^c$ is the effective vertex for production of the
system ${\{f\}}$ (see Fig.~\ref{ImpactFactorRepres}) in the collision of the particle $A$ off the Reggeized gluon with color index $c$ and momentum $-q_1$, with
\begin{equation}
q_1 = \alpha k_2 + {q_1}_{\perp}~,\ \ \ \ \ \ \ \alpha 
\approx -\left( s_{AR} - m_A^2 + \vec q_1^{\:2} \right)/s \ll 1~, 
\end{equation}
and $s_{AR}$ is the particle-Reggeon squared invariant mass. In the fragmentation region of the particle $A$, 
where all transverse momenta as well as the invariant mass $\sqrt{s_{AR}}$ are not 
growing with $s$, we have $q_1^2 = - \vec q_1^{\:2}$.
The factor 
\begin{equation}
    \langle cc^{\prime} | \hat{\cal P}_{0} | 0 \rangle = \frac{\delta^{c c'}}{\sqrt{N^2-1}}
\end{equation}
is the projector on the singlet color state representation. Summation in
eq.~(\ref{ImpactUnpro}) is carried out over all  systems $\{f\}$ which can be
produced in the NLLA and the integration is performed over the phase
space volume of the produced system, which for a $n$-particle
system (if there are identical particles in this system, 
corresponding symmetry factors should also be introduced) reads
\begin{equation}
d\rho_f = (2\pi)^D\delta^{(D)}\biggl(p_A - q_1 - \sum_{m=1}^nk_m\biggr)
\prod_{m=1}^n\frac{d^{D-1}k_m}{2E_m(2\pi)^{D-1}}~,
\label{GenPhasSpa}
\end{equation}
as well as over the particle-Reggeon invariant mass. The average over initial-state color and spin degrees of freedom is implicitly assumed.
The parameter $s_{\Lambda}$, limiting the integration region over the invariant mass in the first term in the R.H.S. of eq.~(\ref{ImpactUnpro}), is introduced for the separation of the contributions of multi-Regge and quasi-multi-Regge kinematics (MRK and QMRK) and should be considered as tending to infinity.
The dependence of the impact factors on this parameter disappears due to the cancellation between the first and the second term in
the R.H.S. of eq.~(\ref{ImpactUnpro}). 
In the second term, usually called ``BFKL counter-term'', $\Phi_{AA}^{(0)}$ is the Born
contribution to the impact factor, which does not depend on $s_0$, while
${\cal K}^{(0)}_r$ is the part of the BFKL kernel in the Born approximation 
connected with real particle production,
\begin{equation}
{\cal K}^{(0)}_r(\vec q_1, \vec q_2) =
\frac{2 g^2 N}{(2\pi)^{D-1}} \frac{1}{(\vec q_1 - \vec q_2)^2}\;.
\label{BornKer}
\end{equation}

\section{Amplitudes}
\label{sec:amplitudes}

In this section, we re-derive the amplitudes needed to construct the impact factor. The latter were first calculated in Ref.~\cite{DelDuca:2001fn}, which we use for comparison. Amplitudes are produced using {\tt FormCalc}~\cite{Hahn:1998yk} and, then, the result is organized as in Ref.~\cite{DelDuca:2001fn}.

\subsection{The off-shell $ggH$ amplitude}
\label{ssec:ggH_vertex}

The off-shell two gluons-Higgs vertex can be expanded on the tensor basis
\beqa
 \left[ g^a(q_1^\mu) g^b(q_2^\nu) H (q_3) \right] &\equiv&  i \delta^{ab} T^{\mu\nu}(q_1, q_2) \nn \\ &=& i \delta^{ab} \left\{  F_T(q_1, q_2)  t_T^{\mu\nu}(q_1, q_2)  + F_L(q_1, q_2)  t_L^{\mu\nu}(q_1, q_2)  \right\}  \nn \;,
\eqa
with
\beqa
t_T^{\mu\nu}(q_1, q_2) &=& q_1 \cdot q_2 \, g^{\mu\nu} - q_2^\mu q_1^\nu \;,\nn \\
t_L^{\mu\nu}(q_1, q_2) &=& q_1^2 \, q_2^2  g^{\mu\nu}  +  q_1 \cdot q_2 \, q_1^\mu q_2^\nu  -  q_2^2 \, q_1^\mu q_1^\nu -  q_1^2  \, q_2^\mu q_2^\nu \;,
\label{Eq:FundamentalTriangularAmo}
\eqa
which can be obtained as solutions of the two Ward-identities $q_{1 \mu} T^{\mu\nu} = q_{2 \nu} T^{\mu\nu} = 0$. 
The fact that the two QED-like Ward identities are valid independently of the on-shellness of the gluons is due to the fact that the production of the Higgs \textit{via} two gluons fusion is essentially Abelian due to the absence of the triple-gluon vertex. The two coefficients $F_T$ and $F_L$  are free from IR and UV divergences. Moreover, they are finite and non-zero for $q_1^2 = 0$ or $q_2^2 = 0$. If one of the two gluons is both on-shell and transverse, the longitudinal structure vanishes
\beqa
\epsilon_\mu(q_1) t_L^{\mu\nu}(q_1, q_2) |_{q_1^2 = 0} = \epsilon_\nu(q_2) t_L^{\mu\nu}(q_1, q_2) |_{q_2^2 = 0} = 0 \, .
\eqa 

\noindent The full expressions for the two coefficients are
\beqa
F_L(q_1,q_2) &=& \frac{\alpha_s m_t^2}{\pi v} \bigg\{  \frac{1}{2 \det \mathcal G} \left\{ \left[ 2 - \frac{3 q_1^2 \, q_2 \cdot p_H}{\det \mathcal G} \right] ( B_0(q_1^2) -  B_0(m_H^2))  \right. \nn \\
&+& \left[ 2 - \frac{3 q_2^2 \, q_1 \cdot p_H}{\det \mathcal G} \right] ( B_0(q_2^2) -  B_0(m_H^2)  )   \nn \\
 &-& \left. \left[ 4 m_t^2 + q_1^2 + q_2^2 + m_H^2 - 3 m_H^2 \frac{q_1^2 q_2^2 }{\det \mathcal G} \right]  C_0(q_1^2, q_2^2) - 2   \right\} \bigg\} \,, 
\label{Eq:FLtri}
\eqa
\beqa
F_T(q_1,q_2) &=& \frac{\alpha_s m_t^2}{\pi v} \bigg\{ \frac{1}{2 \det \mathcal G} \left\{  m_H^2 \left[  B_0(q_1^2) +  B_0(q_2^2)  - 2  B_0(m_H^2) - 2 q_1 \cdot q_2  C_0(q_1^2, q_2^2)   \right] \right. \nn \\
&+& \left. (q_1^2  - q_2^2) ( B_0(q_1^2) -  B_0(q_2^2)) \right\}  \bigg\}    - q_1 \cdot q_2  F_L(q_1^2,q_2^2) \;,
\label{Eq:FTtri}
\eqa
where $p_H = q_1 + q_2$, $\det \mathcal G = q_1^2 q_2^2 - (q_1 \cdot q_2)^2$ is the Gram determinant, $B_0$ and $C_0$ are the bubble and triangle master integrals and $v^2 = 1/(G_F \sqrt{2})$ where $G_F$ is the Fermi constant. In the previous equations we used the compact notation
\beqa
B_0(p^2) \equiv B_0(p^2, m_t^2, m_t^2) \,, \qquad  C_0(p_1^2, p_2^2) \equiv C_0(p_1^2, p_2^2, (p_1 + p_2)^2, m_H^2 ,m_t^2, m_t^2) \;,
\eqa
with
\beqa
B_0(p^2, m_1^2, m_2^2) &=& \frac{\mu^{4-D}}{i \pi^{D/2} r_\Gamma} \int d^D q \frac{1}{[q^2 - m_1^2] [(q+p)^2 - m_2^2]} \;, \\
\eqa
\begin{gather}
    C_0(p_1^2, p_2^2, (p_1+p_2)^2,m_1^2, m_2^2, m_3^2) = \frac{\mu^{4-D}}{i \pi^{D/2} r_\Gamma} \nn \\ \times \int d^D q \frac{1}{[q^2 - m_1^2] [(q+p_1)^2 - m_2^2] [(q+p_1+p_2)^2 - m_3^2]} \;, 
\end{gather}
and
\beqa
r_\Gamma = \frac{\Gamma^2(1-\epsilon) \Gamma(1 + \epsilon)}{\Gamma(1-2\epsilon)}, \qquad D = 4 - 2 \epsilon \; .
\eqa
Bubble integrals contain UV-singularities, but these cancel out immediately in the various combinations. Then, the coefficients $F_T$ and $F_L$ do not exhibit infrared singularities. Moreover, the result agrees with Ref.~\cite{DelDuca:2001fn}. \\
In the large-top-mass limit, $F_T$ provides the leading contribution, while $F_L$ is suppressed by $1/m_t^2$. From explicit computation we get
\beqa
F_T(q_1, q_2) &=&  \frac{\alpha_s}{\pi v}   \left[ -\frac{1}{3} - \frac{7 m_H^2 + 11 q_1^2 + 11 q_2^2}{360 m_t^2} \right]+  \mathcal O (m_t^{-4}) \;,
\nn \\
F_L(q_1, q_2) &=&  \frac{\alpha_s}{180 \pi  v m_t^2}  + \mathcal O (m_t^{-4}) \,.
\eqa
The result is in agreement with the vertex extracted from the effective Lagrangian 
\beqa
\mathcal L_{ggH} = \frac{\alpha_s}{12 \pi} F^{a \, \mu\nu}F^a_{\mu\nu} \frac{H}{v} \,.
\eea

\subsection{The off-shell $gggH$ amplitude}
\label{ssec:gggH_vertex}

The six diagrams contributing to the four-point $gggH$ amplitude are related in pairs by charge conjugation that eventually fixes the color structure of the box by selecting only the antisymmetric structure functions $f^{abc}$. For what concerns the tensor structure, instead, we have the following general decomposition 
\begin{equation}
 \left[ g^a(q_1^\alpha) g^b(q_2^\beta) g^c(q_3^\gamma) H (k_4) \right] \equiv g f^{abc} B^{\alpha\beta\gamma}(q_1, q_2, q_3) = 
g f^{abc} \left[ g^{\alpha \beta} q_{i}^{\gamma} +  q_{i}^{\alpha} q_{j}^{\beta} q_{k}^{\gamma} + {\rm{perm}}. \right] \; .
\end{equation}
Considering that in our case the three gluons will be transverse\footnote{Recall that in the case of the $t$-channel gluon, its effective (\textit{non-sense}) polarization is transverse to its momenta.}, this very general and complicated form can be greatly simplified: 
\beqa
B^{\alpha\beta\gamma}(q_1, q_2, q_3) &=& B_a(q_1,q_2,q_3) \, g^{\alpha \beta} q_1^\gamma  +   B_a(q_2,q_3,q_1) \,    g^{\beta\gamma} q_2^\alpha   + B_a(q_3,q_1,q_2) \,    g^{\alpha\gamma} q_3^\beta     \nn \\
&-&    B_a(q_2,q_1,q_3) \,   g^{\alpha\beta} q_2^\gamma  -   B_a(q_1,q_3,q_2) \,    g^{\alpha\gamma} q_1^\beta         -   B_a(q_3,q_2,q_1) \,    g^{\beta\gamma} q_3^\alpha     \nn \\
&+& B_b(q_1,q_2,q_3) \, q_3^{\alpha} q_3^{\beta} q_1^{\gamma}  + B_b(q_2,q_3,q_1) \,   q_1^{\beta} q_1^{\gamma} q_2^{\alpha}   +   B_b(q_3,q_1,q_2) \,   q_2^{\alpha} q_2^{\gamma} q_3^{\beta} \nn \\
&-& B_b(q_2,q_1,q_3) \,  q_3^{\alpha} q_3^{\beta} q_2^{\gamma}  -  B_b(q_1,q_3,q_2) \, q_2^{\alpha} q_2^{\gamma} q_1^{\beta}    -  B_b(q_3,q_2,q_1) \, q_1^{\beta} q_1^{\gamma} q_3^{\alpha}         \nn \\
&+& B_c(q_1,q_2,q_3)  q_1^{\gamma}  q_2^{\alpha}  q_3^{\beta}    -  B_c(q_2, q_1, q_3)   q_1^{\beta} q_2^{\gamma}  q_3^{\alpha}  \; , \label{Eq:General_box_amplitude}
\eqa
where we also exploited the symmetry of the amplitude under the exchange of any two pairs of gluons. In the expression above, all momenta are taken to be incoming. 
The three independent coefficients in the expression above are explicitly given by
\beqa
B_a(q_1,q_2,q_3)  &=& \frac{\alpha_s m_t^2}{\pi v} \bigg\{ - \frac{1}{2} q_2 \cdot q_3 \left[ D_0(q_1,q_2,q_3) + D_0(q_2,q_3,q_1) + D_0(q_3,q_1,q_2) \right]  \nn \\
&+& q_1 \cdot q_2 \left[  D_{13}(q_2,q_3,q_1) + D_{12}(q_3,q_1,q_2) - D_{13}(q_3, q_2,q_1) \right] \nn \\
&-& 4 \left[ D_{313}(q_2,q_3,q_1)  +  D_{312}(q_3,q_1,q_2)  - D_{313}(q_3,q_2,q_1)   \right] - C_0(q_1, q_2+q_3) \bigg\} \;, \nn \\
B_b(q_1,q_2,q_3)  &=& - \frac{\alpha_s m_t^2}{\pi v} \bigg\{  
D_{13}(q_1,q_2,q_3) + D_{12}(q_2,q_3,q_1) - D_{13}(q_2,q_1,q_3) \nn \\
&+& 4 \left[  D_{37}(q_1,q_2,q_3) + D_{23}(q_1,q_2,q_3)  + D_{38}(q_2,q_3,q_1)   \right. \nn \\
&+& \left. D_{26}(q_2,q_3,q_1) -  D_{39}(q_2,q_1,q_3)  -  D_{23}(q_2,q_1,q_3)     \right]
\bigg\} \;,\nn \\
B_c(q_1,q_2,q_3)  &=& - \frac{\alpha_s m_t^2}{\pi v} \bigg\{
  - \frac{1}{2} \left[ D_0(q_1,q_2,q_3) +  D_0(q_2,q_3,q_1)  + D_0(q_3,q_1,q_2) \right] \nn \\ 
  &+&   4   \left[  D_{26}(q_1, q_2, q_3) + D_{26}(q_2,q_3,q_1) + D_{26}(q_3,q_1,q_2)  \right. \nn \\
  &+& \left.  D_{310}(q_1, q_2, q_3) + D_{310}(q_2,q_3,q_1)  + D_{310}(q_3,q_1,q_2) \right]
 \bigg\} \;,
\eqa
where we employed the notation of~\cite{Passarino:1978jh,DelDuca:2001fn}\footnote{Up to a different sign definition of $C_0, D_{312}$ and $D_{313}$.}. We performed a numerical comparison with the {\tt VBFNLO} code, finding full accord~\cite{Baglio:2011juf,Baglio:2024gyp}; moreover, we checked analytically that, in the on-shell limit, the result agrees with the helicity amplitudes computed in Refs.~\cite{Baur:1989cm,Rozowsky:1997dm}. \\

\noindent Alternatively, we can re-express the results by exploiting the following mapping:
\beqa
\label{eq:D_PV}
D_{12}(q_1, q_2, q_3) &\equiv& D_{2}(\textrm{arg}) + D_{3}(\textrm{arg}) \;,\nn \\
D_{13}(q_1, q_2, q_3) &\equiv& D_{3}(\textrm{arg}) \;,\nn \\
D_{23}(q_1, q_2, q_3) &\equiv& D_{33}(\textrm{arg}) \;,\nn \\
D_{26}(q_1, q_2, q_3) &\equiv& D_{23}(\textrm{arg}) + D_{33}(\textrm{arg}) \;,\nn \\
D_{37}(q_1, q_2, q_3) &\equiv& D_{133}(\textrm{arg}) + D_{233}(\textrm{arg})  + D_{333}(\textrm{arg}) \;, \nn \\
D_{38}(q_1, q_2, q_3) &\equiv& D_{223}(\textrm{arg}) + 2 D_{233}(\textrm{arg})  + D_{333}(\textrm{arg}) \;, \nn \\
D_{39}(q_1, q_2, q_3) &\equiv& D_{233}(\textrm{arg}) + D_{333}(\textrm{arg}) \;, \nn \\
D_{310}(q_1, q_2, q_3) &\equiv& D_{123}(\textrm{arg}) + D_{133}(\textrm{arg}) + D_{223}(\textrm{arg}) + 2 D_{233}(\textrm{arg}) + D_{333}(\textrm{arg}) \;, \nn \\
D_{312}(q_1, q_2, q_3) &\equiv& D_{002}(\textrm{arg}) + D_{003}(\textrm{arg}) \;,\nn \\
D_{313}(q_1, q_2, q_3) &\equiv& D_{003}(\textrm{arg}) \;,
\eqa
with $\textrm{arg} = q_1^2, q_2^2, q_3^2, q_4^2, (q_1+q_2)^2, (q_2+q_3)^2, m_t^2,m_t^2,m_t^2,m_t^2$. \\

The $D$ coefficients on the right-hand sides of eq.~(\ref{eq:D_PV}) are defined from the expansions of the four-point tensor integrals
\beqa
D_{\mu_1 \ldots \mu_n} = \frac{\mu^{4-D}}{i \pi^{D/2} r_\Gamma} \int d^D \ell \frac{\ell_{\mu_1} \ldots \ell_{\mu_n}}{\mathcal D(0, m_0) \mathcal D(p_1, m_1) \mathcal D(p_2, m_2) \mathcal D(p_3, m_3)} \;,
\label{Eq:DTensDefinition}
\eqa
where the denominators are $\mathcal D(p, m) = (\ell + p)^2 - m^2$. In particular, up to rank-3, we have
\beqa
D_\mu &=& \sum_{i = 1}^3 p_{i \mu} D_i \,, \nn \\
D_{\mu\nu} &=& g_{\mu\nu} D_{00} + \sum_{i,j = 1}^3 p_{i, \mu}p_{j, \nu} D_{ij} \,, \nn \\
D_{\mu\nu\rho} &=& \sum_{i=1}^3 \left( g_{\mu\nu} p_{i,\rho} + g_{\nu\rho} p_{i,\mu} + g_{\mu\rho} p_{i,\nu}  \right) D_{00i} + \sum_{i,j,l =1}^3 p_{i, \mu}p_{j, \nu} p_{l, \rho} D_{ijl} \;,
\eqa
with $p_1 = q_1$, $p_2 = q_1 + q_2$, $p_3 = q_1 + q_2 + q_3$ and $p_4 = q_1 + q_2 + q_3 + q_4$.\footnote{The main difference between the notation used here and in \cite{DelDuca:2001fn} is in the momenta employed in the tensor expansion. Here we used the ones appearing in the denominators, $p_i$, while in \cite{DelDuca:2001fn} the external momenta $q_i$ have been employed.} \\
In the large-top-mass limit the three coefficients are given by
\beqa
B_a(q_1,q_2,q_3) &=& \frac{\alpha_s}{\pi v} \bigg[ \frac{1}{3} +  \frac{1}{360 m_t^2}  
( 11 q_1^2 + 40 q_2^2 + 27 q_3^2 - 4 q_4^2 + 11 (q_1 + q_3)^2 \nn  \\
&-& 14 (q_2 + q_3)^2 )
 \bigg]   + \mathcal O (m_t^{-4}) \;, \nn \\
B_b(q_1,q_2,q_3) &=& \frac{11 \alpha_s }{180 \pi v m_t^2}  + \mathcal O (m_t^{-4}) \;,\nn \\
B_c(q_1,q_2,q_3) &=& \frac{ \alpha_s }{5 \pi v m_t^2}  + \mathcal O (m_t^{-4}) \;,
\eqa
with $B_a$ providing the leading contribution. \\

\section{Kinematics and LO computation}
\label{sec:kinematicsLO}

\subsection{Kinematics}
\label{ssec:kinematics}

It is useful to decompose any four-vector into the Sudakov basis formed by the two light-like vectors, $k_1$ and $k_2$, with $k_1 \cdot k_2 = s/2$. Within this basis, we can introduce the Sudakov decompositions 
\bea
&& p_H = z_H \, k_1 + \frac{m_H^2 + \vec{p}_H^{\; 2}}{z_H s} k_2 + p_{H_\perp} \,, \qquad  p_{H_\perp}^2 = - \vec{p}_H^{\; 2}\,, \qquad  p_H^2 = m_H^2 \;, \nn \\
&& q = - \alpha k_2 + q_\perp \,, \qquad q^{\; 2} = q_\perp^2 = - \vec{q}^{\; 2} \;, \nn \\
&& p_p = z_p \, k_1 + \frac{{\vec{p}_p}^{\;2}}{z_p s} k_2 + p_{p_\perp} \,, \qquad  p_{p_\perp}^{2} = - \vec{p}_p^{\; 2} \,, \qquad  p_p^2 = 0 \; ,
\eea
where $p_H$ is the Higgs momentum, $q$ the $t$-channel Reggeon momenta and $p_p$ the momentum of the additional particle produced in the NLO case ($p=q$ for the quark and $p=g$ for the gluon). The relevant scalar products among the momenta can be expressed in terms of the variables $z_H,z_q, \vec{q}, \vec{p}_H, \vec{p}_q$. The invariant mass of the particle-Reggeon system, in the case of real corrections, can be easily expressed as
\begin{equation}
   s_{AR} = (k_1+q)^2 = (p_p + p_H)^2 = \frac{z_p (z_H + z_p) m_H^2 + (z_p \vec{p}_H   -    z_H   \vec{p}_p)^2}{z_H z_p} \; 
\end{equation}
and, therefore, the integration measure of the NLO impact factor reads
\begin{equation}
\frac{d s_{pR}}{2 \pi} d \rho_{ \{H q \}} = \delta(1 - z_p - z_H) \ \delta^{(2)}(\vec{p}_p + \vec{p}_H - \vec{q}) \ \frac{d z_p d z_H}{z_p z_H}\  \frac{d^{D-2}p_p \, d^{D-2} p_H}{2 (2\pi)^{D-1}} .  
\label{Eq:PhasePhaseInt}
\end{equation}
Since we want to obtain a impact factor differential in the Higgs kinematic variables, we will only perform the integration over the longitudinal fraction and the transverse momentum of the additional produced particle by using the Dirac delta functions appearing in eq.~(\ref{Eq:PhasePhaseInt}). For this reason, in calculating effective vertices, we can immediately use the constraints
\begin{equation}
    z_p = 1- z_H \; , \hspace{3 cm} \vec{p}_p = \vec{q} - \vec{p}_H \; .
\end{equation}

\subsection{LO computation}
\label{ssec:LO computation}

\begin{figure}
\begin{picture}(430,60)
\put(120,35){ \scalebox{5.0}{( }}
\put(150,17){\includegraphics[scale=0.27]{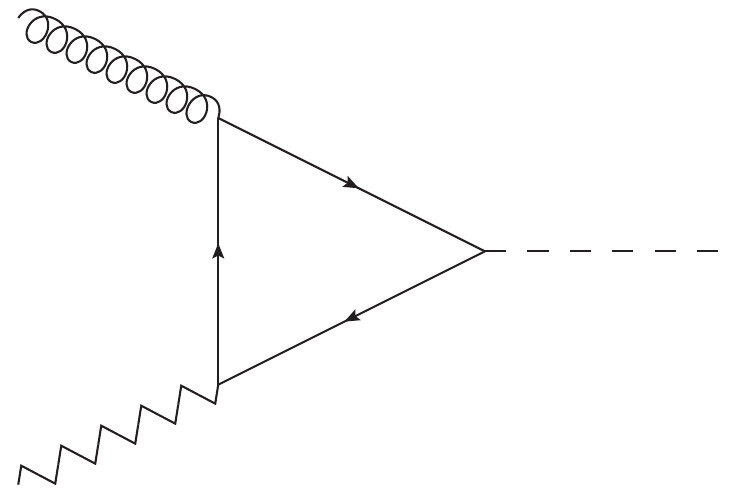}}
\put(168,80){\scalebox{0.8}{$k_1$}}
\put(168,16){\scalebox{0.8}{$q$}}
\put(222,56){\scalebox{0.8}{$p_H$}}
\put(193,0){ \textbf{t-channel} }
\put(264,47){ $\times \hspace{0.15 cm} 2$}
\put(270,35){ \scalebox{5.0}{ ) }}
\end{picture} 
\caption{The two triangular-like diagrams contributing to the Higgs impact factor at the LO. The $\times 2$ indicates the diagram in which the direction of the fermionic lines is reversed. The momenta of the initial state particles (the collinear parton and the Reggeized gluon) are considered to be incoming, while the momenta of the final state particles (the Higgs in the present case) are taken outgoing.}
\label{Fig:2DiagramsTriLO}
\end{figure}
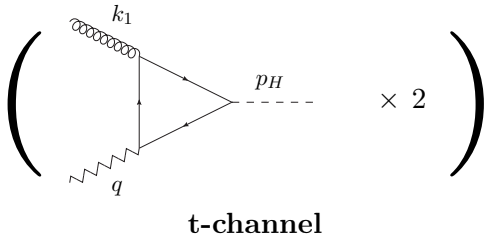

The LO impact factor for the production of a Higgs in the gluon-Reggeon collision reads
\begin{equation}
\Phi_{gg}^{ \{ H \} }(\vec q) = \frac{\langle cc^{\prime} | \hat{\cal P}_0 | 0 \rangle }{2(1-\epsilon) (N^2-1)}
\sum_{a, \lambda } \int \frac{ds_{AR}}{2\pi}\ d\rho_f \ \Gamma_{\{ H \}g}^{ac} (\vec q \; )
\left( \Gamma_{\{ H \}g}^{ac^{\prime}} (\vec q \; ) \right)^* \; ,
\label{Eq:LoImpactD4-2E}
\end{equation}
where we average over color and polarizations of the incoming gluon in $D=4-2\epsilon$. The effective vertex of interaction entering eq.~(\ref{Eq:LoImpactD4-2E}) reads
\begin{equation}
  \Gamma_{\{ H \}g}^{ac} (\vec q \; ) =  (-i) i \delta^{ac} T^{\mu \nu} (k_1,q) \varepsilon_{\perp , \mu }(k_1) \left( -\frac{k_{2, \nu }}{s} \right) = \frac{F_T ( k_1 , q ) (q_{\perp} \cdot \varepsilon_{\perp} (k_1) ) \delta^{ac}}{2} \; ,
  \label{EffectivegRH}
\end{equation}
where $T^{\mu \nu}$  is the tensor defined in eq.~(\ref{Eq:FundamentalTriangularAmo}) and $F_T$ the function in~(\ref{Eq:FTtri}) and for the physical gluon we employ the gauge choice $\varepsilon (k_1) \cdot k_2 = 0$. This same choice will be used for all physical gluons in the NLO calculation.  We observe that, in general, $F_T$ can be expressed as a function of $k_1^2, q^2=-\vec{q}^{\; 2}$ and $(k_1+q) = p_H^2 = m_H^2$, {\it i.e.}\footnote{The dependence on $m_t^2$ is understood.}
\begin{equation}
    F_T (k_1 , q) \equiv  F_T \left(k_1^2, q^{2}, (k_1+q)^2 \right) = F_T \left(0, -\vec{q}^{\; 2}, m_{H}^2  \right)  \; ,
\end{equation}
where we have used the on-shell condition of the collinear gluon, $k_1^2=0$.
In the infinite-top-mass limit, the effective vertex in eq.~(\ref{EffectivegRH}) reduces to
\begin{equation}
  \Gamma_{\{ H \}g}^{ac} (\vec q \; ) = - \frac{g_H (q_{\perp} \cdot \varepsilon_{\perp} (k_1) ) \delta^{ac}}{2} \; ,
  \label{EffectivegRHMtInfinity}
\end{equation}
in agreement with Ref.~\cite{Celiberto:2022fgx}. \\

\noindent The gluon-initiated impact factor differential in the Higgs kinematic variables reads
\begin{equation}
  \frac{ d \Phi_{gg}^{ \{ H \} (0) }(\vec q) }{d z_H d^{2} \vec{p}_H } = \frac{| F_T \left( 0, -\vec{q}^{\; 2}, m_{H}^2  \right) |^2 \vec{q}^{\; 2} }{8 (1-\epsilon) \sqrt{N^2-1}}  \delta (1-z_H) \delta^{(2)} \left( \vec{q} -\vec{p}_H \right) \; .
  \label{Eq:LoImpactD4-2EPart}
\end{equation}
The convolution with the gluon parton distribution function allows us to define the proton-initiated impact factor as
\begin{equation}
  \frac{ d \Phi_{PP}^{ \{ H \} (0) }(\vec q \; ) }{d x_H d^{2} \vec{p}_H } \!=\! \int_{x_H}^1 \hspace{-0.2 cm} \frac{d z_H}{z_H} f_g \hspace{-0.1 cm}  \left( \hspace{-0.05 cm}  \frac{x_H}{z_H} \hspace{-0.05 cm}  \right)  \frac{ d \Phi_{gg}^{ \{ H \} (0) }(\vec q \; ) }{d z_H d^{2} \vec{p}_H } = \frac{ |F_T \left( 0, -\vec{q}^{\; 2}, m_{H}^2  \right)|^2 \vec{q}^{\; 2} f_g (x_H) }{8 (1-\epsilon) \sqrt{N^2-1}} \delta^{(2)} \left( \vec{q} -\vec{p}_H \right),
  \label{Eq:LoImpactD4-2EHadro}
\end{equation} 
which, in the infinite-top-mass limit, gives us the result already obtained in Ref.~\cite{Celiberto:2022fgx}.

\section{NLO computation}
\label{sec:NLOcomputation}

\subsection{Quark-initiated contribution}
\label{ssec:NLOquark}

\begin{figure}
\begin{picture}(430,60)
\put(120,35){ \scalebox{5.0}{( }}
\put(150,20){\includegraphics[scale=0.36]{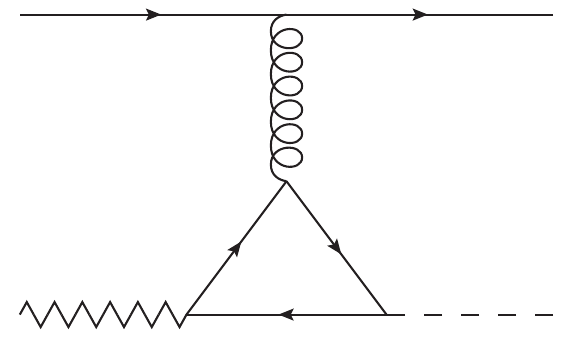}}
\put(165,85){\scalebox{0.8}{$k_1$}}
\put(224,85){\scalebox{0.8}{$p_q$}}
\put(165,15){\scalebox{0.8}{$q$}}
\put(224,15){\scalebox{0.8}{$p_H$}}
\put(193,0){ \textbf{t-channel} }
\put(264,47){ $\times \hspace{0.15 cm} 2$}
\put(270,35){ \scalebox{5.0}{ ) }}
\end{picture} 
\caption{The two triangular-like diagrams contributing to the quark-initiated contribution to the Higgs impact factor at the NLO. Again the $\times 2$ indicates the diagram in which the direction of the fermionic lines is reversed. The momenta of the initial state particles (the collinear parton and the Reggeized gluon) are considered to be incoming, while the momenta of the final state particles (the Higgs and the outgoing quark) are taken outgoing.}
\label{Fig:2DiagramsTriQ}
\end{figure}

For the process initiated by a quark, the impact factor is given by
\bea
d \Phi_{qq}^{\{H q\}} = \langle cc' | \hat{\mathcal P}_0 | 0 \rangle \frac{1}{2 N} \sum_{i,j, \lambda, \lambda'} \int \frac{ds_{qR}}{2\pi} d \rho_{\{Hq\}} \Gamma^{c(0)}_{\{Hq\}q}(q) \left( \Gamma^{c'(0)}_{\{Hq\}q}(q) \right)^* \;,
\eea
where the average (sum) over the spin states and the color configurations of the quark of the incoming (outgoing) fermion has been already carried out. The vertex $\Gamma^{c(0)}_{\{Hq\}q}$ is computed from the diagrams depicted in Fig.~\ref{Fig:2DiagramsTriQ}, using the results given in Sec.~\ref{ssec:ggH_vertex}. The final expression reads
\bea
\Gamma^{c(0)}_{\{Hq\}q}(q) \! &=& \! - g t^c_{ij}  \frac{ 1-z_H}{(\vec{p}_H - \vec{q})^2} \bar u(p_q) \bigg\{  F_T((p_H - q)^2, q^2, m_H^2) \bigg[   (p_H - q) \cdot q  \frac{\hat{k}_2}{s}  -  (p_H - q)\cdot \frac{k_2}{s} \hat{q} \bigg]      \nn \\
&-&  F_L((p_H - q)^2, q^2, m_H^2)  \frac{(\vec{p}_H - \vec{q})^2}{ 1-z_H} q^2 \frac{\hat{k}_2}{s}   \bigg\} u(k_1) \;,
\eea
(here $\hat a \equiv \gamma^\mu a_\mu$) from which one can easily evaluate the differential impact factor
\bea
&& \frac{d \Phi_{qq}^{\{Hq\}} (z_H, \vec{p}_H, \vec{q} ) }{d z_H d^2 \vec{p}_H} = 
 \frac{g^2  \sqrt{N^2-1}}{16 z_H N (2 \pi)^{D-1} (\vec{q}-\vec{p}_H )^4}   \nn \\
&& \times \bigg\{  |F_T((p_H - q)^2, q^2, m_H^2)|^2   [4 (1-z_H) [ ( \vec{q}  -  \vec{p}_H )\cdot \vec{q} \; ]^2 + z_H^2 \vec{q}^{\; 2}  ( \vec{q}  -  \vec{p}_H )^2  ]    \nn \\
&&+ 4 \Re\{ F_T((p_H - q)^2, q^2, m_H^2)^* F_L((p_H - q)^2, q^2, m_H^2)\} (2 - z_H) ( \vec{q}  -  \vec{p}_H )^2 \vec{q}^{\; 2}   \,   ( \vec{q}  -  \vec{p}_H ) \cdot \vec{q}   \nn \\
&&+ 4 |F_L((p_H - q)^2, q^2, m_H^2)|^2  ( \vec{q} - \vec{p}_H )^4 \vec{q}^{\; 4}  \bigg\}  \,.
\eea 
As expected from gauge invariance, the vertex and the impact factor vanish for $\vec{q} \to \vec{0}$. A quick comparison with the computation performed in the infinite-top-mass limit~\cite{Celiberto:2022fgx} highlights the presence of a genuinely new contribution, given by the longitudinal form factor $F_L$, when we retain the full top mass dependence. At leading order in the heavy top expansion, we recover the result of~\cite{Celiberto:2022fgx}. The impact factor contains a collinear singularity when $\vec{p}_q = \vec{q} - \vec{p}_H \rightarrow \vec{0}$, which is fully contained in the first term of curly brackets; thus, as expected, the infrared structure of the impact factor is not influenced by the details of the coupling between the real gluon and the Reggeon to produce the Higgs (and therefore is completely analogous to the one found by exploiting the infinite-top-mass approximation~\cite{Celiberto:2022fgx}). 

\subsection{Gluon-initiated contribution}
\label{ssec:NLOgluon}

\subsubsection{$gR \rightarrow gH$ \textit{via} a triangular-like contribution}
\label{sssec:triangle}

\begin{figure}
\begin{picture}(430,150)
\put(30,115){\includegraphics[scale=0.36]{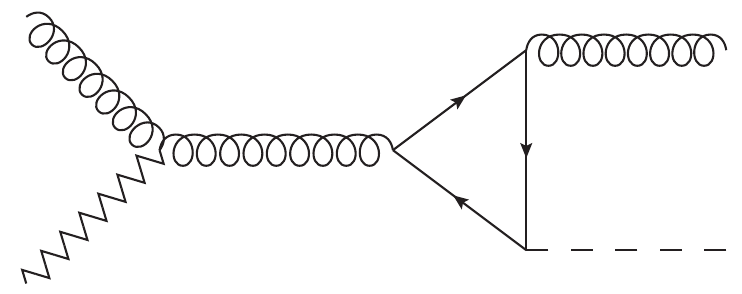}}
\put(5,127){ \scalebox{5.0}{( }}
\put(164,137){ $\times \hspace{0.15 cm} 2$}
\put(50,155){\scalebox{0.8}{$k_1$}}
\put(50,125){\scalebox{0.8}{$q$}}
\put(137,165){\scalebox{0.8}{$p_g$}}
\put(137,115){\scalebox{0.8}{$p_H$}}
\put(160,127){ \scalebox{5.0}{ ) }}
\put(203,135){ \scalebox{1.5}{ + }}
\put(230,127){ \scalebox{5.0}{( }}
\put(275,175){\scalebox{0.8}{$k_1$}}
\put(347,175){\scalebox{0.8}{$p_H$}}
\put(263,110){\includegraphics[scale=0.36]{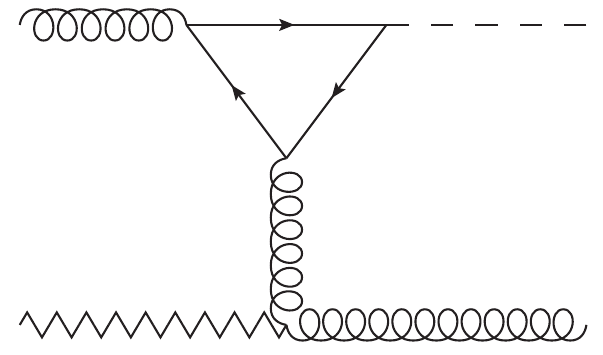}}
\put(275,105){\scalebox{0.8}{$q$}}
\put(347,105){\scalebox{0.8}{$p_g$}}
\put(370,137){ $\times \hspace{0.15 cm} 2$}
\put(375,127){ \scalebox{5.0}{ ) }}
\put(80,92){ \textbf{s-channel} }
\put(298,92){ \textbf{u-channel} }
\put(120,35){ \scalebox{5.0}{( }}
\put(150,20){\includegraphics[scale=0.36]{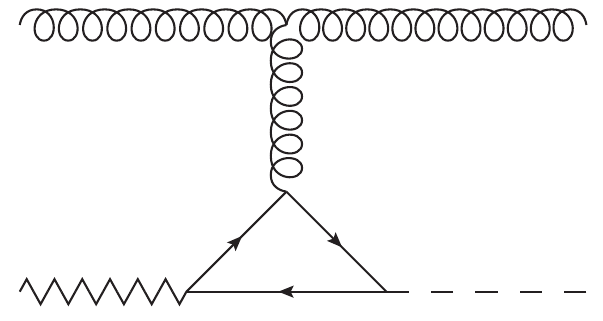}}
\put(165,81){\scalebox{0.8}{$k_1$}}
\put(224,81){\scalebox{0.8}{$p_g$}}
\put(165,15){\scalebox{0.8}{$q$}}
\put(224,15){\scalebox{0.8}{$p_H$}}
\put(193,0){ \textbf{t-channel} }
\put(264,47){ $\times \hspace{0.15 cm} 2$}
\put(270,35){ \scalebox{5.0}{ ) }}
\put(85,45){ \scalebox{1.5}{ + }}
\end{picture} 
\caption{The six triangular-like diagrams contributing to the gluon-initiated contribution to the Higgs impact factor at the NLO. Again the $\times 2$ indicates the diagram in which the direction of the fermionic lines is reversed. The momenta of the initial state particles (the collinear parton and the Reggeized gluon) are considered to be incoming, while the momenta of the final state particles (the Higgs and the outgoing quark) are taken outgoing.}
\label{Fig:6DiagramsTri}
\end{figure}
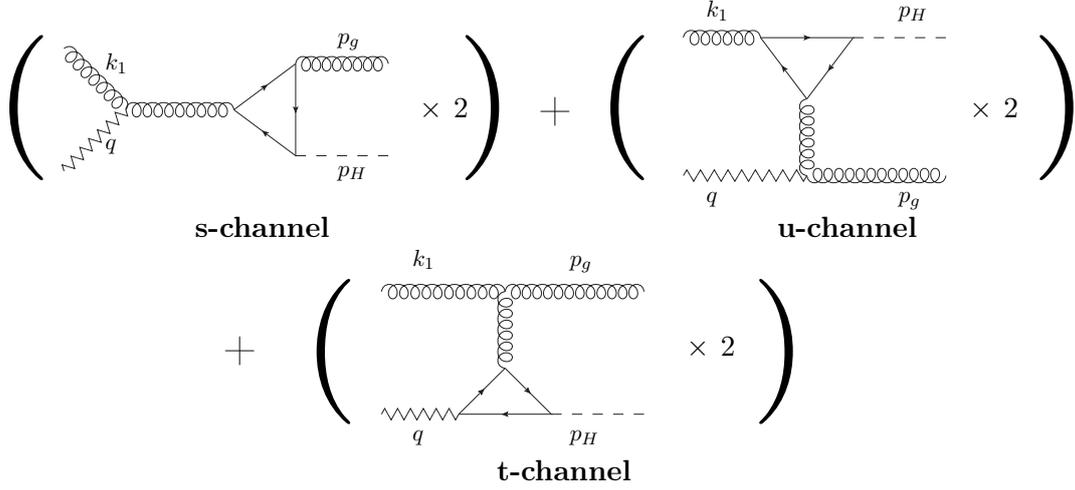

In this section we inspect the role of the $ggH$ vertex in the $gR \to g H$ amplitudes. The Feynman diagrams in Fig.~\ref{Fig:6DiagramsTri} can be further labeled as $s,t$- and $u$-channel diagrams. In the $s$- and $u$-channel diagrams, one of the two gluon lines emerging from the triangle loop (respectively, $p_g$ and $k_1$) is on-shell and transverse. As such, the longitudinal form factor does not contribute, as shown in Sec.~\ref{ssec:ggH_vertex}. In this case the tensor structure of the $ggH$ vertex is the same as for the heavy-top limit and the comparison with the results presented in \cite{Celiberto:2022fgx} is straightforward. For the $t$-channel diagrams, instead, both gluon lines attached to the quark loops are off-shell (even though the \textit{effective} polarization vector of the Reggeized gluon satisfies the transversality condition $\displaystyle \frac{k_2}{s} \cdot q = 0$) and the $ggH$ vertex also contributes with 
\bea
F_L(p^2, q^2, (p+q)^2) \left[ p^2 \, q^2  g^{\rho\nu}    -  q^2 \, p^\rho p^\nu    \right] \frac{k_{2,\nu}}{s} \;,
\eea
where $p = k_1 - p_g = p_H - q$. \\
The Feynman diagrams shown in Fig.~\ref{Fig:6DiagramsTri} give
\bea
\Gamma_\triangle = \Gamma_s + \Gamma_t + \Gamma_u \;,
\eea
with 
\begin{gather}
\Gamma_s = i \frac{g f^{abc} F_T((p_g + p_H)^2, 0, m_H^2) }{2 [ m_H^2 (1-z_H) +  (\vec{p}_H - z_H \vec{q})^2 ]} \left[ 
(m_H^2 (1- z_H)^2 + (\vec{p}_H - z_H \vec{q})^2 )   g^{\mu\nu}   \right. \nn \\
+ \left. 2 z_H (1 - z_H)^2  p_H^{\mu} p_H^{\nu}
- 2 z_H^2 (1 - z_H)  p_g^{\mu} p_H^{\nu}
 \right] \epsilon_{\mu}(k_1)\epsilon^*_{\nu}(p_g) \;, 
\end{gather}
 \begin{gather}
 \Gamma_u = -i \frac{g f^{abc} F_T(0, (p_H - k_1)^2, m_H^2) }{2 [  m_H^2 (z_H - 1) -   \vec{p}_H^{\; 2} ]} \nn \\ \times \left[
  (m_H^2 + \vec{p}_H^{\; 2}) (1 - z_H) g^{\mu\nu} - 2 z_H p_H^{\mu} (p_H^{\nu} - z_H k_1^{\nu})
 \right] \epsilon_{\mu}(k_1)\epsilon^*_{\nu}(p_g) \; ,
 \end{gather} 
 \begin{gather}
  \Gamma_t = i  \frac{g f^{abc} F_T((k_1- p_g)^2, q^2, m_H^2) }{(\vec{q} - \vec{p}_H)^2} (1 \hspace{-0.05 cm} - \hspace{-0.05 cm} z_H) \hspace{-0.05 cm} \left[
  (\vec{q} - \vec{p}_H) \hspace{-0.05 cm}  \cdot \hspace{-0.05 cm}  \vec{q} g^{\mu\nu} \hspace{-0.1 cm} + \hspace{-0.1 cm}  z_H (p_H^{\mu} k_1^{\nu} + p_g^{\mu} p_H^{\nu})
  \right] \epsilon_{\mu}(k_1)\epsilon^*_{\nu}(p_g)  \nn \\
  - i \frac{g f^{abc} F_L((k_1- p_g)^2, q^2, m_H^2)}{2} (z_H - 2) \vec{q}^{\; 2}  \epsilon(k_1) \cdot \epsilon^*(p_g) \;,
\end{gather}
which are in agreement with the ones in~\cite{Celiberto:2022fgx} in the infinite-top-mass limit $F_T(p^2, q^2) = - \frac{\alpha_s}{ 3 \pi v} $ and $F_L = 0$. By expressing the amplitudes in terms of the transverse degrees of freedom, we finally find 
\begin{gather}
\Gamma_s = i \frac{g f^{abc} F_T ( (p_g + p_H)^2, 0, m_H^2) }{2 [ m_H^2 (1-z_H) +  (\vec{p}_H - z_H \vec{q})^2 ]} \left[ 
(m_H^2 (1- z_H)^2 + (\vec{p}_H - z_H \vec{q})^2 )   g^{\mu\nu}   \right. \nonumber \\
+ \left. 2 z_H (p_{\perp H} - z_H q_\perp)^{\mu} (p_{\perp H} - z_H q_\perp)^{\nu}
 \right] \epsilon_{\perp \mu}(k_1)\epsilon^*_{\perp \nu}(p_g) \; , 
\label{Eq:Gammas}
\end{gather}
\begin{gather}
 \Gamma_u = -i \frac{g f^{abc} F_T (0, (p_H - k_1)^2, m_H^2) }{2 [  m_H^2 (z_H - 1) -   \vec{p}_H^{\; 2} ]} \\ \times \left[
  (m_H^2 + \vec{p}_H^{\; 2}) (1 - z_H) g^{\mu\nu} - 2 z_H p_{\perp H}^{\mu} p_{\perp H}^{\nu} 
 \right] \epsilon_{\perp \mu}(k_1) \epsilon^*_{\perp \nu}(p_g) \; , 
 \label{Eq:Gammau}
\end{gather}
\begin{gather}
   \Gamma_t = i  \frac{g f^{abc} F_T ((k_1- p_g)^2, q^2, m_H^2)  }{(\vec{q} - \vec{p}_H)^2}  \left[
  (1-z_H) (\vec{q} - \vec{p}_H) \cdot \vec{q} g^{\mu\nu} 
  \hspace{-0.1 cm} +\hspace{-0.1 cm}  z_H  q_{\perp}^{\mu}  p_{\perp H}^{\nu} 
  \hspace{-0.05 cm}  - \hspace{-0.1 cm}  z_H ( 1 - z_H) p_{\perp H}^{\mu} q_{\perp}^{\nu} \right. \nonumber \\ 
  - \left. z_H^2 q_{\perp}^{\mu} q_{\perp}^{\nu}
  \right] \epsilon_{\perp \mu}(k_1)\epsilon^*_{\perp \nu}(p_g) 
  - i \frac{g f^{abc} F_L ((k_1- p_g)^2, q^2, m_H^2)  }{2} (z_H - 2) \vec{q}^{\; 2}  \epsilon_{\perp}(k_1) \cdot \epsilon_{\perp}^*(p_g)   \;.
  \label{Eq:Gammat}
\end{gather}
To obtain the infinite-top-mass result, it is enough to set 
\begin{equation}
 F_T ( (p_g + p_H)^2, 0, m_H^2)  = F_T  (0, (p_H - k_1)^2, m_H^2) = F_T((k_1- p_g)^2, q^2, m_H^2)  = - \frac{\alpha_s}{3 \pi v}  \equiv - g_H 
\end{equation}
and
\begin{equation}
    F_L ((k_1- p_g)^2, q^2, m_H^2) = 0 \; 
\end{equation}
in eqs.~(\ref{Eq:Gammas}), (\ref{Eq:Gammau}) and (\ref{Eq:Gammat}). The result agrees with Ref.~\cite{Celiberto:2022fgx}.

\subsubsection{$gR \rightarrow gH$ \textit{via} a box-like contribution}
\label{sssec:box}

\begin{figure}
\begin{picture}(430,150)
\put(45,105){\includegraphics[scale=0.27]{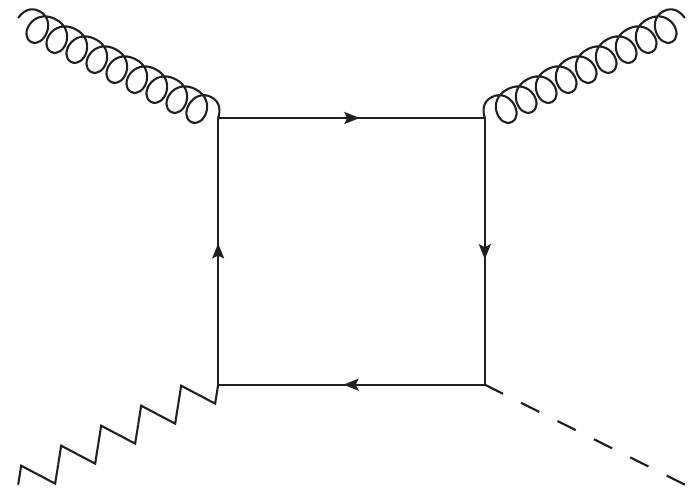}}
\put(58,169){\scalebox{0.8}{$k_1$}}
\put(58,105){\scalebox{0.8}{$q$}}
\put(117,169){\scalebox{0.8}{$p_g$}}
\put(117,105){\scalebox{0.8}{$p_H$}}
\put(5,127){ \scalebox{5.0}{( }}
\put(154,137){ $\times \hspace{0.15 cm} 2$}
\put(160,127){ \scalebox{5.0}{ ) }}
\put(203,135){ \scalebox{1.5}{ + }}
\put(230,127){ \scalebox{5.0}{( }}
\put(263,110){\includegraphics[scale=0.27]{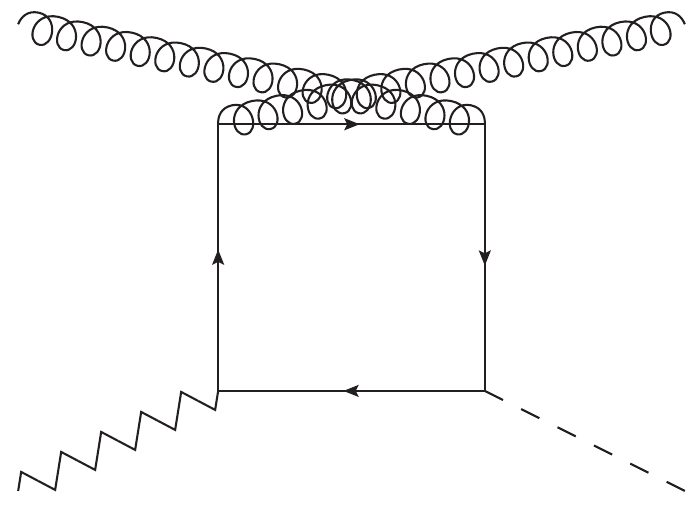}}
\put(278,175){\scalebox{0.8}{$k_1$}}
\put(278,110){\scalebox{0.8}{$q$}}
\put(327,175){\scalebox{0.8}{$p_g$}}
\put(327,110){\scalebox{0.8}{$p_H$}}
\put(370,137){ $\times \hspace{0.15 cm} 2$}
\put(375,127){ \scalebox{5.0}{ ) }}
\put(120,35){ \scalebox{5.0}{( }}
\put(150,20){\includegraphics[scale=0.27]{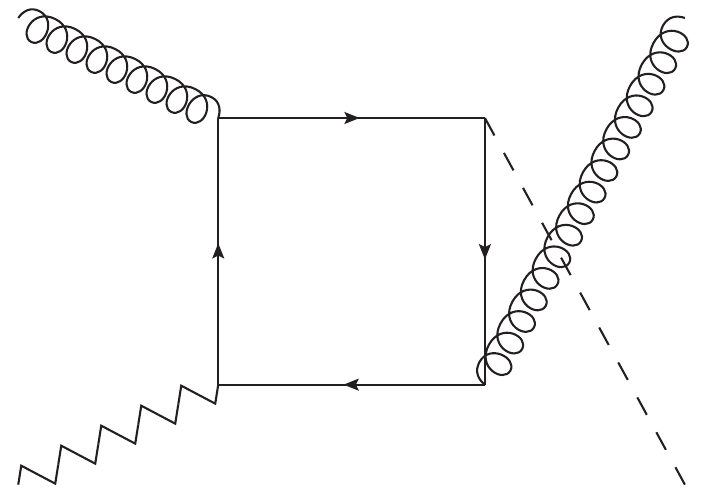}}
\put(168,78){\scalebox{0.8}{$k_1$}}
\put(168,20){\scalebox{0.8}{$q$}}
\put(222,78){\scalebox{0.8}{$p_g$}}
\put(222,20){\scalebox{0.8}{$p_H$}}
\put(264,47){ $\times \hspace{0.15 cm} 2$}
\put(270,35){ \scalebox{5.0}{ ) }}
\put(85,45){ \scalebox{1.5}{ + }}
\end{picture} 
\caption{The six box-like diagrams contributing to the gluon initiated contribution to the Higgs impact factor at the NLO. Again the $\times 2$ indicates the diagram in which the direction of the fermionic lines is reversed. The momenta of the initial state particles (the collinear parton and the Reggeized gluon) are considered to be incoming, while the momenta of the final state particles (the Higgs and the outgoing quark) are taken outgoing.}
\label{Fig:6DiagramsBox}
\end{figure}

Using eq.~(\ref{Eq:General_box_amplitude}) and expressing the result in terms of transverse polarization vectors, we find
\beqa
\Gamma_\Box &=& -i  \left[ g^a(k_1^\mu) g^c(q^\rho) g^b(-p_g^\nu) H (k_4) \right] \nn \\
&=& -i g f^{abc}   \frac{1}{2} \bigg\{ -\left[ B_a(k_1, -p_g, q) + (1 - z_H) B_a(-p_g, k_1, q)  \right] g^{\mu\nu}     \nn \\
 &-& [B_b(k_1, -p_g, q)  + (1 - z_H) B_b(-p_g, k_1, q)]  q_\perp^{\mu} q_\perp^{\nu}  \nn \\
&-&   (1 - z_H) B_b(q, k_1, -p_g) r_\perp^{\mu} q_\perp^{\nu}  - \frac{1}{1 - z_H} B_b(q, -p_g, k_1) q_\perp^{\mu} r_\perp^{\nu} \nn \\
&-&  \left[ B_b(k_1, q, -p_g)  + \frac{1}{1 - z_H}  B_b(-p_g, q, k_1) \right] r_\perp^{\mu} r_\perp^{\nu} 
    \nn \\
&+&   B_c(q, k_1, -p_g)  r_\perp^{\mu} q_\perp^\nu   +  B_c(k_1, q, -p_g) q_\perp^\mu    r_\perp^\nu       \bigg\} \epsilon_{\perp \mu}(k_1) \epsilon^*_{\perp \nu}(p_g) \;,
\label{Eq:Box_Amplit}
\eqa
with $r_\perp^\mu=q_\perp^\mu- p_{\perp H}^\mu$. In the large-top-mass limit $\Gamma_{\Box}$ becomes
\beqa
 \Gamma_\Box \big |_{m_t \rightarrow \infty} = i g g_H f^{abc}  \frac{2 - z_H}{2}  \epsilon_{\perp}(k_1) \cdot \epsilon^*_{\perp}(p_g) \;,
\eqa
which exactly reproduces the results obtained in \cite{Celiberto:2022fgx}.  

\subsubsection{Infrared structure and check of gauge invariance}
\label{sssec:infraredgauge}

As expected, the real contribution to the impact factor contains infrared phase-space singularities. The soft ones will be canceled in the combination between real and virtual corrections, while a remaining divergence of a purely collinear nature will be removed when the renormalization of the gluon PDF is carried out. The relevant infrared singularities can be classified as
\begin{itemize}
    \item[\textbullet] \textbf{Collinear singularity}: when $\vec{p}_g = \vec{q}-\vec{p}_H \rightarrow \vec{0}$ and $ z_g = 1- z_H$ is fixed, 
    \item[\textbullet] \textbf{Soft singularity}: after parameterizing $\vec{p}_g = \vec{q}-\vec{p}_H = (1-z_H) \vec{u}$, $z_g= 1- z_H \rightarrow 0$ .
\end{itemize}
We do not include to this list, the rapidity divergence characteristic of the BFKL approach:
\begin{itemize}
    \item[\textbullet] \textbf{Rapidity singularity}: $z_H \rightarrow 1$ and $(\vec{q}-\vec{p}_H)$ is fixed,
\end{itemize}
which is removed by the BFKL counter-term in eq.~(\ref{ImpactUnpro}) and to which we reserve the entire Section~\ref{sssec:Rapiditydivergence}. We stress that this latter limit is the only one in which the Higgs-gluon invariant mass goes to infinity corresponding to a wide separation in rapidity. \\

\noindent In the present case, the infrared divergences classified above are contained entirely in the interference term resulting from the multiplication of the $F_T$-part of the $\Gamma_t$-term with its complex conjugate. This latter contribution is proportional to the square of
\begin{equation}
    F_T( k_1- p_g , q) =  F_T((k_1- p_g)^2 , -\vec{q}^{\; 2} , m_H^2) \xrightarrow[\text{collinear}]{\text{soft and/or}} F_T(0, -\vec{p}_H^{\; 2} , m_H^2)  \; ,
\end{equation}
which coincides with the analogous structure present in the LO impact factor. Once this structure is factorized, it is easy to observe that the infrared-singular part of the contribution is simply a generalization of that obtained in the infinite-top-mass limit (see last line of eq.~(3.17) of~\cite{Celiberto:2022fgx}), where $g_H^2$ is replaced by $|F_T(0, -\vec{p}_H^{\; 2}, m_H^2)|^2$. The fact that the soft and collinear divergences come from the $t$-channel diagram is not a surprise. Indeed, the expected collinear divergence is associated with the initial state radiation of the collinear gluon. Even in the soft case, this is the only diagram in which the additional gluon is emitted from an external on-shell gluon line. In the other diagrams, the soft limit is ``protected'' either by the top mass or by the virtuality of the $t$-channel Reggeon. \\

\noindent Given the complexity of the final result, it is useful to extract its infrared singularities explicitly. The complete result for the infrared limit of the impact factor is
\begin{gather}
\frac{d \Phi_{gg}^{ \{ Hg \} } \left( z_H, \vec{p}_H, \vec{q}; s_0 \right)}{d z_H d^2 \vec{p}_H}  \xrightarrow{\vec{q} \; \sim \; \vec{p}_H} \frac{g^2 \left| F_T \left(0, -\vec{p}_H^{\; 2}, m_H^2 \right) \right|^2 N}{4 (1-\epsilon) \sqrt{N^2-1} (2 \pi)^{D-1}} \frac{\vec{p}_H^{\; 2}}{(\vec{q}- \vec{p}_H)^{2}} \frac{z_H}{1 - z_H} \theta (s_{\Lambda}-s_{gR}) \nonumber \\ + \frac{g^2 \left| F_T \left(0, -\vec{p}_H^{\; 2}, m_H^2 \right) \right|^2 N}{4 (1-\epsilon) \sqrt{N^2-1} (2 \pi)^{D-1}} \frac{1}{(\vec{q} - \vec{p}_H)^2} \left[ z_H (1-z_H) \vec{p}_H^{\; 2} + 2 (1-\epsilon) \frac{1-z_H}{z_H} \frac{(\vec{p}_H \cdot (\vec{q}- \vec{p}_H))^2}{(\vec{q} - \vec{p}_H)^2} \right] . 
\label{Eq:IFinTheCollLim}
\end{gather}
The first term contains a soft divergence, while the second and third contain collinear divergences. It is important to observe that, the first divergence is of soft nature and it is not a rapidity divergence. Indeed, it should be noted that, in a cut-off regularization of the singularities related to the longitudinal fraction $z_g$, terms that go with the inverse of $z_g$ are associated with both rapidity and soft divergences. However, in the present case, in the high-rapidity limit, $z_H \rightarrow 1$ at fixed $\vec{q}- \vec{p}_H$, the $t$-channel diagrams do not even contribute to the rapidity divergence of the impact factor, due to a $(1-z_H)$-suppressing factor generated by the structure $F_T$ (see Section \ref{sssec:Rapiditydivergence}). For better clarity, we note that, while in the collinear/soft limit
\begin{gather}
    F_T((k_1- p_g)^2 , -\vec{q}^{\; 2} , m_H^2) = F_T \left( - \frac{(\vec{q} - \vec{p}_H)^2}{1-z_H} , -\vec{q}^{\; 2} , m_H^2 \right) \xrightarrow[\text{collinear}]{\text{soft and/or}} F_T(0, -\vec{p}_H^{\; 2} , m_H^2) \; ,
\end{gather}
in the rapidity limit
\begin{gather}
    F_T((k_1- p_g)^2 , -\vec{q}^{\; 2} , m_H^2) \nonumber  \\ = F_T \left( - \frac{(\vec{q} - \vec{p}_H)^2}{1-z_H} , -\vec{q}^{\; 2} , m_H^2 \right)  \xrightarrow{\text{rapidity}} F_T( \infty, -\vec{p}_H^{\; 2} , m_H^2) \sim (1-z_H) \; .
\end{gather}
Thus, the divergence in the first term of eq.~(\ref{ImpactUnpro}) is soft. If one considers the infrared approximation of the impact factor, the dependence on the parameter $s_{\Lambda}$ disappears in combination with the counter-term BFKL (second term of eq.~(\ref{ImpactUnpro})), which, taken in the same approximation, is also soft divergent. As already explained, we want to add and subtract a term that makes the impact factor completely finite (removing both rapidity and infrared divergences). Anticipating the result of Section \ref{sssec:Rapiditydivergence} for the rapidity divergence, we can construct a subtraction term,
\begin{gather}
\frac{d \Phi_{gg}^{ \{ Hg \} } \left( z_H, \vec{p}_H, \vec{q}; s_0 \right)}{d z_H d^2 \vec{p}_H} \bigg |_{\rm div.} \hspace{-0.5 cm} = \frac{g^2 \left| F_T \left(0, -\vec{p}_H^{\; 2}, m_H^2 \right)  \right|^2 N}{4 (1-\epsilon) \sqrt{N^2-1} (2 \pi)^{D-1}} \frac{\vec{q}^{\; 2}}{(\vec{q}- \vec{p}_H)^{2}} \frac{z_H}{1 - z_H} \theta \left(s_{\Lambda} - \frac{(\vec{q}-\vec{p}_H)^2}{1-z_H} \right) \nonumber \\ + \frac{g^2 \left| F_T \left(0, -\vec{p}_H^{\; 2}, m_H^2 \right)  \right|^2 N}{4 (1-\epsilon) \sqrt{N^2-1} (2 \pi)^{D-1}} \frac{1}{(\vec{q} - \vec{p}_H)^2} \left[ z_H (1-z_H) \vec{q}^{\; 2} + 2 (1-\epsilon) \frac{1-z_H}{z_H} \frac{(\vec{q} \cdot (\vec{q}- \vec{p}_H))^2}{(\vec{q} - \vec{p}_H)^2} \right] \; ,
\label{Eq:IFCombinedSub}
\end{gather}
capable of completely removing both the rapidity and IR divergences in the impact factor. We want to stress that, although terms in eqs.~(\ref{Eq:IFinTheCollLim}) and~(\ref{Eq:IFCombinedSub}) are equivalent in the collinear limit $q \rightarrow p_H$\footnote{The $\theta$-function is needed only in the region $z_H \rightarrow 1$.}, the rapidity divergence in~(\ref{Eq:IFCombinedSub}) does not come from diagrams of $\Gamma_t$. \\

A stringent cross-check of the result is the proof of gauge invariance. Indeed, the definition of impact factor employed here is explicitly gauge invariant and this requires that, for $\vec{q} = \vec{0}$, the full amplitude vanishes. Therefore, in this limit, the triangular contribution, $\Gamma_{\triangle}$, should exactly cancel the box contribution, $\Gamma_{\Box}$. In order to show explicitly the gauge invariance of the calculation, namely the cancellation of the different contributions for $\vec{q} \to \vec{0}$, we exploit the large-top-mass expansion. Up to the next-to-leading order, the triangular amplitude reads
\begin{equation}
\begin{split}
& \Gamma_{\triangle}|_{\vec{q} = 0} =  -i  g f^{abc} \frac{\alpha_s}{3 \pi v}   \frac{2- z_H}{2}  \epsilon_{\perp}(k_1) \cdot \epsilon^*_{\perp}(p_g) + i  g f^{abc} \left( -\frac{\alpha_s }{360 \pi v m_t^2}  \right) \nn \\
& \times \frac{2-z_H}{2(1 - z_H)} \left[  ( 11 \vec{p}_H^{\; 2} + 7 m_H^2 (1-z_H))   g^{\mu\nu}  + 22  p_{\perp H}^{\mu} p_{\perp H}^{\nu} \right] \epsilon_{\perp \mu}(k_1) \epsilon^*_{\perp \nu}(p_g) + \mathcal O(m_t^{-4}) \nn \; ,
\end{split}
\end{equation}
while the box amplitude gives
\begin{equation}
\begin{split}
& \Gamma_{\Box}|_{\vec{q} = 0} = i  g f^{abc} \frac{\alpha_s}{3 \pi v}   \frac{2- z_H}{2}  \epsilon_{\perp}(k_1) \cdot \epsilon^*_{\perp}(p_g) -i g f^{abc}  \left( -\frac{\alpha_s }{360 \pi v m_t^2}  \right) \nn \\ & \times \frac{2 - z_H}{2(1 - z_H)}  \left[  (11 \vec{p}_H^{\; 2} + 7 m_H^2 (1 - z_H))  g^{\mu\nu} + 22 p_{H \perp}^{\mu} p_{H \perp}^{\nu} \right]  \epsilon_{\perp \mu}(k_1) \epsilon^*_{\perp \nu}(p_g) + \mathcal O(m_t^{-4}) \, .
\end{split}    
\end{equation}
The sum of the two terms exactly cancels, order by order, as expected. Using {\tt Package X}, we have verified this statement up to terms of order $m_t^{-4}$ (next-to-next-to-leading order).

\subsubsection{The gluon-initiated impact factor}
\label{sssec:gluonIF}

The total amplitude of the $gR \to gH$ process can be represented in a compact way as
\[
\Gamma^{abc(0)}_{\{Hg\}g}(q) = i g f^{abc}  \left[ C_{00} g_{\mu\nu} + C_{11} p_{\perp H}^\mu p_{\perp H}^\nu + C_{12} p_{\perp H}^\mu q_{\perp}^\nu \right.
\]
\beq
\left.
+ C_{21} q_{\perp}^\mu p_{\perp H}^\nu + C_{22} q_{\perp }^\mu q_{\perp}^\nu \right]  \epsilon_{\perp \mu}(k_1) \epsilon^*_{\perp \nu}(p_g) \;,
\eeq
where the coefficients $C_{ij}$ are given by
\begin{gather}
C_{00} =   \frac{m_H^2 (1-z_H)^2 +  (\vec{p}_H - z_H \vec{q})^2}{2[ m_H^2 (1-z_H) +  (\vec{p}_H - z_H \vec{q})^2 ]} F_T((p_g + p_H)^2, 0, m_H^2) \nn \\
- \frac{(1 - z_H) (m_H^2 + \vec{p}_H^{\,2} )}{2 [ m_H^2 (z_H - 1) -   \vec{p}_H^{\,2}] } F_T(0, (p_H - k_1)^2, m_H^2) 
+ \! (1-z_H) \frac{(\vec{q} - \vec{p}_H) \cdot \vec{q} }{(\vec{q} - \vec{p}_H)^2} F_T((k_1- p_g)^2, q^2, m_H^2)  \nn \\
+ \frac{ (2-z_H) \vec{q}^{\,2} }{2}   F_L((k_1- p_g)^2, q^2, m_H^2) + \frac{1}{2} \left[ B_a(k_1, -p_g, q) + (1 - z_H) B_a(-p_g, k_1, q)  \right]  \; , 
\end{gather}
\begin{gather}
    C_{11} = z_H \frac{F_T((p_g + p_H)^2, 0, m_H^2)}{ m_H^2 (1-z_H) +  (\vec{p}_H - z_H \vec{q})^2 }  + z_H \frac{F_T(0, (p_H - k_1)^2, m_H^2)}{m_H^2 (z_H - 1) -   \vec{p}_H^2} \nn \\
+ \frac{1}{2} \left[ B_b(k_1, q, -p_g)  + \frac{1}{1 - z_H}  B_b(-p_g, q, k_1) \right]  \; , 
\end{gather}
\begin{gather}
  C_{12} = - z_H^2 \frac{F_T((p_g + p_H)^2, 0, m_H^2)}{ m_H^2 (1-z_H) +  (\vec{p}_H - z_H \vec{q})^2 }  - \frac{z_H (1 - z_H)}{(\vec{q} - \vec{p}_H)^2} F_T((k_1- p_g)^2, q^2, m_H^2)  \nn \\
- \frac{1}{2}   \left[    B_b(k_1, q, -p_g) +  \frac{1}{ 1 - z_H}   B_b(-p_g, q, k_1) + (1 - z_H) B_b(q, k_1, -p_g)     -  B_c(q, k_1, -p_g) \right] \; ,   
\end{gather}
\begin{gather}
    C_{21} = - z_H^2 \frac{F_T((p_g + p_H)^2, 0, m_H^2)}{ m_H^2 (1-z_H) +  (\vec{p}_H - z_H \vec{q})^2 }  + \frac{z_H }{(\vec{q} - \vec{p}_H)^2} F_T((k_1- p_g)^2, q^2, m_H^2) \nn \\
- \frac{1}{2} \left[ B_b(k_1, q, -p_g)  +  \frac{1}{1 - z_H}B_b(-p_g, q, k_1)  +  \frac{1}{1 - z_H} B_b(q, -p_g, k_1)    -    B_c(k_1, q, -p_g)   \right]  \; ,
\end{gather}
\bea
C_{22} &=&  z_H^3 \frac{F_T((p_g + p_H)^2, 0, m_H^2)}{ m_H^2 (1-z_H) +  (\vec{p}_H - z_H \vec{q})^2 }  -  \frac{z_H^2 }{(\vec{q} - \vec{p}_H)^2} F_T((k_1- p_g)^2, q^2, m_H^2)  \nn \\
&+&  \frac{1}{2}  \bigg[  B_b(k_1, -p_g, q) + B_b(k_1, q, -p_g) + (1 - z_H) B_b(-p_g, k_1, q) + (1 - z_H) B_b(q, k_1, -p_g) \nn \\
&+&  \frac{1}{1 - z_H}  (B_b(-p_g, q, k_1) + B_b(q, -p_g, k_1)) - 
   B_c(k_1, q, -p_g) - B_c(q, k_1, -p_g)   \bigg] \,.
\eea
The sum over the transverse polarizations of the gluons is performed with
\bea
\sum_\lambda \epsilon^{\mu}_{\perp, \lambda}(k)\epsilon^{\nu *}_{\perp, \lambda}(k) = - g^{\mu\nu}_{\perp \perp} = - g^{\mu\nu} + \frac{k_1^{\mu} k_2^{\nu}+ k_2^{\mu} k_1^{\nu}}{k_1 \cdot k_2} \;,
\eea
where $g^{\mu\nu}_{\perp \perp}$ represents the metric tensor in the transverse space. We also take the convolution with the gluon PDF to get the proton-initiated impact factor:
\begin{gather}
\frac{d \Phi_{PP}^{\{H g\}} (x_H, \vec{p}_H, \vec{q}) }{  d z_H d^2 p_H} \!=   \frac{g^2 N  }{ 4 (2\pi)^{D-1} (1-\epsilon) \sqrt{N^2-1} } \int_{x_H}^1 \frac{d z_H}{z_H^2} f_g \left( \frac{x_H}{z_H} \right) \frac{1}{(1-z_H)} \bigg \{ 2 (1-\epsilon) |C_{00}|^2  \nonumber \\ + \vec{p}_H^{\; 4} |C_{11}|^2 +  \vec{q}^{\; 4} |C_{22}|^2 + 2 (\vec{p}_H \cdot \vec{q})^2   \Re\{ C_{12}^* C_{21} + C_{11}^* C_{22} \} 
+ \vec{p}_H^{\; 2} \vec{q}^{\; 2} (|C_{12}|^2 + |C_{21}|^2) \nonumber \\ + 2 \vec{p}_H^{\; 2} \, \vec{p}_H \cdot \vec{q} \, \Re \{C_{11}^* (C_{12} + C_{21}) \} + 2\vec{q}^{\; 2} \, \vec{p}_H \cdot \vec{q} \, \Re \{C_{22}^* (C_{12} + C_{21}) \} -  2 \vec{p}_H^{\; 2} \, \Re \{ C_{00}^* C_{11} \} \nonumber  \\
 - 2 \vec{q}^{\; 2} \, \Re \{ C_{00}^* C_{22} \} - 2 \vec{p}_H \cdot \vec{q} \, \Re \{C_{00}^* (C_{12} + C_{21}) \}  \bigg\} \; \; \theta \left( s_{\Lambda} - \frac{(1-z_H) m_H^2 + \vec{\Delta}^2}{z_H (1-z_H)} \right) \,.
\label{eq:gRgH_square}
\end{gather}
The contribution from a real gluon emission has also a divergence for $z_H \to 1$ ($ z_g \to 0$) at any value of the outgoing gluon transverse momenta $\vec{q}-\vec{p}_H$. This is the rapidity divergence and it is regulated by the parameter $s_{\Lambda}$. In the final result, it should be cancelled by the BFKL counter-term appearing in the definition of the NLO impact factor. Unfortunately, unlike the infrared case, the treatment of this (high-energy) singularity is much more complex and it does not follow the same path as the infinite-top-mass limit case. \\

\subsubsection{Rapidity divergence}
\label{sssec:Rapiditydivergence}
In Ref.~\cite{DelDuca:2003ba}, the high-energy factorization for Higgs-plus-two-jet production has been analyzed in two distinct limits: (a) the Higgs boson centrally located in rapidity between the two jets, and very far from either; (b) the Higgs boson close in rapidity to one \textit{identified} jet, and both of these very far from the other jet. The second allows to extract the impact factor for the production of a Higgs in association with a jet. The crucial difference between the abovementioned case and the present calculation is that, since in our case we are more inclusive in the final state, we can explore the region where $z_H \rightarrow 1$ ($z_g \rightarrow 0$), which is affected by the rapidity divergence associated with the gluon-Higgs invariant mass going to infinity. This singularity is just a consequence of the separation of contributions associated with MRK and QMRK kinematics~\cite{Fadin:1998fv} and, in our definition of the impact factors, eq.~(\ref{ImpactUnpro}), cancels in the combination between the first (QMRK) and the second (MRK) term. To demonstrate and make explicit this cancellation, however, it is necessary to study the limit in which Higgs and additional gluon are highly separated in rapidity. Considering again the paradigmatic example of the Higgs-plus-two-jet production, this corresponds to the kinematic configuration (c) in which the Higgs boson is emitted forwardly and is well separated from the centrally emitted jet, the latter being strongly separated in rapidity from the other backward jet. In this case, the amplitude should take a Regge form with the central gluon emission described by the Lipatov vertex. \\

Thus, in order to extract from our result the rapidity divergence and demonstrate that it corresponds to the one predicted by the BFKL factorization and cancelled by the BFKL counter-term ({\it i.e.} the last term in eq.~(\ref{ImpactUnpro})), we need to expand around $z_H = 1$ the contributions in eqs.~(\ref{Eq:Gammas}), (\ref{Eq:Gammau}), (\ref{Eq:Gammat}), (\ref{Eq:Box_Amplit}). It is interesting to compare the cases of infinite and finite top mass. In the former case~\cite{Celiberto:2022fgx}, $F_T = -B_a = -g_H$, while $F_L = B_b = B_c = 0$. Due to the constant behavior of these structures and to the phase-space factor in eq.~(\ref{Eq:PhasePhaseInt}), all diagrams contribute to the $z_H \rightarrow 1$ limit. However, thanks to a huge simplification, the impact factor in the $z_H \rightarrow 1$ limit coincides with the contribution given, in the same limit, by just the $\Gamma_t$ term. This is equivalent to saying that, in $z_H \rightarrow 1$ limit, the contributions from all other diagrams cancel each other out. \\

In the finite-top-mass case, the structures $F_T, F_L, B_a, B_b$ and $B_c$ are non-trivial functions of $z_H$ and their asymptotic behavior in the $z_H \rightarrow 1$ must be carefully studied to determine the rapidity limit of the impact factor. Let us start our discussion from the triangular functions $F_T ((p_g+p_H)^2, 0, m_H^2), F_T (0, (p_H-k_1)^2, m_H^2), F_T((k_1-p_g)^2, 0, m_H^2)$ and $F_L ((k_1-p_g)^2, 0, m_H^2)$.
The asymptotic behaviors of the ``external masses'' entering these functions are
\begin{gather}
(p_g+p_H)^2 \xrightarrow{z_H \sim 1} \frac{(\vec{q} - \vec{p}_H)^2}{1-z_H} \; , \hspace{1 cm} (k_1-p_g)^2  \xrightarrow{z_H \sim 1} - \frac{(\vec{q} - \vec{p}_H)^2}{1-z_H} \nonumber \\ 
(p_H-k_1)^2 \xrightarrow{z_H \sim 1} - \vec{p}_H^{\; 2} \; , \hspace{1 cm}  p_g^2 = k_1^2 = 0 \; , \hspace{1 cm} p_H^2 = m_H^2 \; , \hspace{1 cm} q^2 = -\vec{q}^{\; 2}  \;.
\end{gather}
From the definitions of $F_L$ and $F_T$ (eqs.~(\ref{Eq:FLtri}) and~(\ref{Eq:FTtri}), respectively), it is easy to see that they scale with the inverse of their external masses. For this reason, the diagrams of $\Gamma_t$ and $\Gamma_s$ undergo a linear suppression in $(1-z_H)$ which ensures that they do not contribute to the high-rapidity limit. Then, the whole $z_H \to 1$ contribution of the triangular diagrams comes from $\Gamma_u$, which does not contains any suppressing scale. The fact that diagrams, which are finite in the high-energy limit, become divergent when the infinite-top-mass approximation is taken in the first instance, is a manifestation of the non-commutativity of the limits $s \rightarrow \infty$ and $m_t \rightarrow \infty$.  \\

In the case of box-type contributions, the analytic extraction of the $z_H \to 1$ limit is very complicated and we have resorted to the symbolic aid of {\tt Package X}. The LoopRefineSeries function of {\tt Package X}, supported by the standard Series function of {\tt Mathematica}, is able to expand all the structures contributing to the impact factor, with the exception of the box-type scalar integrals
\begin{gather}
  D_0 (0, 0, m_H^2, q^2, -s, s, m_t^2, m_t^2, m_t^2, m_t^2) \; , \nonumber \\
  D_0 (0, 0, m_H^2, q^2, -s, u, m_t^2, m_t^2, m_t^2, m_t^2) \; , \hspace{1 cm} D_0 (0, 0, m_H^2, q^2, s, u, m_t^2, m_t^2, m_t^2, m_t^2) \; ,
\label{EQ:variousBox}
\end{gather}
with
%which are defined, according to {\tt Package X} notation, as in eq.~(\ref{Eq:DTensDefinition}). After the Feynman parameterization, the generic 4-point scalar integral reads
%\begin{gather}
%  D_0(s_1, s_2, s_3, s_4, s_{12}, s_{23}, m_0^2, m_1^2, m_2^2, m_3^2) = \lim_{\varepsilon \rightarrow 0^+} \int_0^1 d x \int_0^{1-x} d y \int_0^{1-x-y} dz \nonumber \\
%  \times \left[ s_1 \; x^2 + s_{12} \; y^2 + s_4 \; z^2 + (s_1 + s_{12} - s_2) xy + (s_1 -s_{23} + s_4) xz + (s_{12} - s_3 + s_4)yz \right. \nonumber \\
%  \left. + ( -s_1 + m_1^2 - m_0^2) x + (-s_{12} + m_2^2 - m_0^2 ) y + ( - s_4 + m_3^2 - m_0^2) z + m_0^2 - i \varepsilon \right]^{-2} \; ,
%\end{gather}
\begin{gather}
   D_0(p_1^2, p_2^2, p_3^2, p_4^2, (p_1+p_2)^2, (p_2+p_3)^2, m_0^2, m_1^2, m_2^2, m_3^2) = \frac{\mu^{4-D}}{i \pi^{D/2} r_\Gamma} \nn \\ \times \int d^D q \frac{1}{[q^2 - m_0^2] [(q+p_1)^2 - m_1^2] [(q+p_1+p_2)^2 - m_2^2] [(q+p_1+p_2+p_3)^2 - m_3^2]}  
\end{gather}
and
\begin{equation}
s = (k_1 + q)^2 \xrightarrow{z_H \sim 1} \frac{(\vec{q} - \vec{p}_H)^2}{1-z_H} \; , \hspace{1 cm} u = (k_1 - p_H)^2 \xrightarrow{z_H \sim 1} - \vec{p}_H^{\; 2} \; .
\label{sAndu}
\end{equation}
For the last two terms in (\ref{EQ:variousBox}), it is enough to observe that they scale as the inverse of $s$ (or linearly in $(1-z_H)$) and therefore do not contribute. Contrarily, the first one must be calculated in the $s \rightarrow \infty$ limit and gives
\begin{gather}
  D_0 (0, 0, m_H^2, q^2, -s, s, m_t^2, m_t^2, m_t^2, m_t^2) = - \frac{1}{s^2} \left[ \left( \ln \left( \frac{s}{m_t^2} \right) - i \pi \right)^2 + \ln^2 \left( \frac{s}{m_t^2} \right) \right. \nonumber \\
 \left. - \ln^2 \left( - \frac{1 + \sqrt{ 1 + 4 m_t^2 / \vec{q_{}}^{\; 2} } }{1 - \sqrt{ 1 + 4 m_t^2 / \vec{q_{}}^{\; 2}  }} \right) -  \ln^2 \left( - \frac{ 1 + \sqrt{   1 - 4 m_t^2 / m_H^{2}  } }{ 1 - \sqrt{ 1 - 4 m_t^2 / m_H^{2} }} \right) \right] \; .
 \label{BoxIntegralAsymptotic}
\end{gather} 
From eqs.~(\ref{sAndu}) and (\ref{BoxIntegralAsymptotic}) is immediate to observe that, in addition to entering as a power, $s \sim 1/(1-z_H)$ appears also in the argument of logarithms. In the box contributions, this occurs in dominant terms of the $z_H \rightarrow 1$ expansion, leading to rapidity divergences of the form
\begin{equation}
\frac{1}{1-z_H} \; , \hspace{1 cm} \frac{1}{1-z_H} \ln (1-z_H) \; , \hspace{1 cm} \frac{1}{1-z_H} \ln^2 (1-z_H) \; .
\label{Rapidity divergences}
\end{equation}
The last two types of terms are incompatible with the ones predicted by the BFKL factorization and indeed cancel individually for all ten coefficients in eq.~(\ref{Eq:Box_Amplit}). Armed with the results given above and using {\tt Package X}, we can final compute the high-rapidity limit of the impact factor, which reads
\begin{equation*}
   \frac{d \Phi_{gg}^{ \{H g \}} (z_H, \vec{p}_H, \vec{q} ; s_0)}{d z_H d^2 \vec{p}_H} \bigg|_{z_H \rightarrow 1} 
\end{equation*}
\begin{equation*}
   = \frac{\langle cc'|\hat{\mathcal{P}}|0 \rangle }{2(1-\epsilon)(N^2-1)} 
    \left[ \sum_{\{ f \}} \int \frac{d s_{PR} d\rho_f}{2 \pi} \Gamma^c_{P \{ f \}} \left( \Gamma^c_{P \{ f \}} \right)^{*} \theta \left( s_{\Lambda} - s_{PR} \right) \right]_{z_H \rightarrow 1}
\end{equation*}
\begin{equation}
    = \frac{g^2 |F_T ( 0, -\vec{p}_H^{\; 2}, m_{H}^2  )|^2 N}{4 (1-\epsilon) \sqrt{N^2-1} (2 \pi)^{D-1} } \frac{\vec{q}^{\; 2}}{(\vec{q}- \vec{p}_H)^2} \frac{1}{(1-z_H)} \theta \left( s_{\Lambda} - \frac{(\vec{q}-\vec{p}_H)^2}{(1-z_H)} \right) \; .
\label{Eq:HighRapidityLimitIF}
\end{equation}
It is worth stress that, although the expression in eq.~(\ref{Eq:HighRapidityLimitIF}) appears as a trivial extension of the infinite-top-mass limit case~\cite{Celiberto:2022fgx}, the mechanisms that led to this result are intrinsically different. Now, we can write eq.~(\ref{Eq:IFCombinedSub}) as
\begin{gather}
\frac{d \Phi_{gg}^{ \{ Hg \} } \left( z_H, \vec{p}_H, \vec{q}; s_0 \right)}{d z_H d^2 \vec{p}_H} \bigg |_{\rm div.} \hspace{-0.5 cm} = z_H \frac{d \Phi_{gg}^{ \{H g \}} (z_H, \vec{p}_H, \vec{q} ; s_0)}{d z_H d^2 \vec{p}_H} \bigg|_{z_H \rightarrow 1}  \nonumber \\ + \frac{g^2 \left| F_T ( 0, -\vec{p}_H^{\; 2}, m_{H}^2  ) \right|^2 N}{4 (1-\epsilon) \sqrt{N^2-1} (2 \pi)^{D-1}} \left[ z_H (1-z_H) \vec{q}^{\; 2} + 2 (1-\epsilon) \frac{1-z_H}{z_H} \frac{(\vec{q} \cdot (\vec{q}- \vec{p}_H))^2}{(\vec{q} - \vec{p}_H)^2} \right] \; .
\end{gather}
After the convolution with the gluon PDF, the subtraction term becomes
\begin{gather}
 \int_{x_H}^1 \frac{d z_H}{z_H} f_g \left( \frac{x_H}{z_H} \right) \frac{d \Phi_{gg}^{ \{ Hg \} } \left( z_H, \vec{p}_H, \vec{q}; s_0 \right)}{d z_H d^2 \vec{p}_H} \bigg |_{\rm div.} \hspace{-0.3 cm} = \int_{x_H}^1 d z_H f_g \left( \frac{x_H}{z_H} \right)  \frac{d \Phi_{gg}^{ \{ H g \}}(z_H, \vec{p}_H, \vec{q};s_0)}{d z_H d^2 \vec{p}_H} \Bigg |_{z_H=1} \nonumber \\ 
 + \int_{x_H}^1 \frac{d z_H}{z_H} f_g \left( \frac{x_H}{z_H} \right) \frac{g^2 \left| F_T ( 0, -\vec{p}_H^{\; 2}, m_{H}^2  ) \right|^2 N}{4 (1-\epsilon) \sqrt{N^2-1} (2 \pi)^{D-1}} \nonumber \\ \times \left[ z_H (1-z_H) \vec{q}^{\; 2} + 2 (1-\epsilon) \frac{1-z_H}{z_H} \frac{(\vec{q} \cdot (\vec{q}- \vec{p}_H))^2}{(\vec{q} - \vec{p}_H)^2} \right] \nonumber \\
 \equiv \int_{x_H}^1 d z_H f_g \left( \frac{x_H}{z_H} \right)  \frac{d \Phi_{gg}^{ \{ H g \}}(z_H, \vec{p}_H, \vec{q};s_0)}{d z_H d^2 \vec{p}_H} \Bigg |_{z_H=1} + \frac{d \Phi^{\{Hg\}{\rm no \; plus}}_{PP}(x_H, \vec{p}_H,\vec q;s_0)}{d x_H d^2 p_H} \;.
\end{gather}
We now combine the BFKL counter-term and the gluon-initiated contribution in a convenient way that makes the cancellation of the rapidity divergence immediate and isolate the IR-singular sector. First of all, we rewrite~(\ref{eq:gRgH_square}) in an equivalent form, by adding and subtracting the following three terms: 
\begin{gather*}
\int_{x_H}^1 \frac{d z_H}{z_H} f_g \left( \frac{x_H}{z_H} \right) \frac{d \Phi_{gg}^{ \{ Hg \} } \left( z_H, \vec{p}_H, \vec{q}; s_0 \right)}{d z_H d^2 \vec{p}_H} \bigg |_{\rm div.} \; , \\ \int_{x_H}^1 d z_H f_g (x_H)  \frac{d \Phi_{gg}^{ \{H g \}} (z_H, \vec{p}_H, \vec{q} ; s_0)}{d z_H d^2 \vec{p}_H} \bigg|_{z_H \rightarrow 1} 
\end{gather*}
and
\begin{gather*}
\int^{x_H}_0 d z_H f_g (x_H)  \frac{d \Phi_{gg}^{ \{H g \}} (z_H, \vec{p}_H, \vec{q} ; s_0)}{d z_H d^2 \vec{p}_H} \bigg|_{z_H \rightarrow 1} \; .
\end{gather*}
Then, we get
\[
\frac{d \Phi^{\{Hg\}}_{PP}(x_H, \vec{p}_H,\vec q;s_0)}{d x_H d^2 p_H}
= \frac{d \tilde{\Phi}^{\{Hg\}}_{PP}(x_H, \vec{p}_H,n,\vec q;s_0)}{d x_H d^2 p_H}
\]
\[
+ \frac{d \Phi^{\{Hg\} (1-x_H)}_{PP}(x_H, \vec{p}_H,\vec q;s_0)}{d x_H d^2 p_H}
+ \frac{d \Phi^{\{Hg\}{{\rm real} \; P_{gg}}}_{PP}(x_H, \vec{p}_H,\vec q;s_0)}{d x_H d^2 p_H}
\]
\begin{equation}
+ \int_{x_H}^1 d z_H f_g (x_H)  \frac{d \Phi_{gg}^{ \{H g \}} (z_H, \vec{p}_H, \vec{q} ; s_0)}{d z_H d^2 \vec{p}_H} \bigg|_{z_H \rightarrow 1} \; ,
\label{subtractions}
\end{equation}
where
\begin{equation*}
    \frac{d \tilde{\Phi}^{\{Hg\}}_{PP}(x_H, \vec{p}_H,\vec q;s_0)}{d x_H d^2 p_H}
    =\int_{x_H}^1 \frac{d z_H}{z_H} f_g \left( \frac{x_H}{z_H} \right) \frac{d \Phi_{gg}^{ \{ H g \}}(z_H, \vec{p}_H, \vec{q};s_0)}{d z_H d^2 \vec{p}_H} 
\end{equation*}
\begin{equation}
    - \int_{x_H}^1 d z_H f_g \left( \frac{x_H}{z_H} \right)  \frac{d \Phi_{gg}^{ \{ H g \}}(z_H, \vec{p}_H, \vec{q};s_0)}{d z_H d^2 \vec{p}_H} \Bigg |_{z_H=1} -  \frac{d \Phi^{\{Hg\}{\rm no \; plus}}_{PP}(x_H, \vec{p}_H,\vec q;s_0)}{d x_H d^2 p_H} \; ,
\label{Eq:TermDiff}
\end{equation}
\begin{equation}
\frac{d \Phi^{\{Hg\} (1-x_H)}_{PP}(x_H, \vec{p}_H,\vec q;s_0)}{d x_H d^2 p_H}
=\int_{0}^{x_H} d z_H f_g (x_H) \frac{d \Phi_{gg}^{ \{ H g \}}(z_H, \vec{p}_H, \vec{q};s_0)}{d z_H d^2 \vec{p}_H} \Bigg |_{z_H=1} 
\label{Eq:Term(1-zH)}
\end{equation}
and
\begin{equation*}
\frac{d \Phi^{\{Hg\}{\rm real} \; P_{gg}}_{PP}(x_H, \vec{p}_H,\vec q;s_0)}{d x_H d^2 p_H}
= - \int_{0}^{x_H} d z_H f_g (x_H) \frac{d \Phi_{gg}^{ \{ H g \}}(z_H, \vec{p}_H, \vec{q};s_0)}{d z_H d^2 \vec{p}_H} \Bigg |_{z_H=1}
\end{equation*}
\begin{equation}
    + \int_{x_H}^1 d z_H \left( f_g \left( \frac{x_H}{z_H} \right) - f_g (x_H) \right) \frac{d \Phi_{gg}^{ \{ H g \}}(z_H, \vec{p}_H, \vec{q};s_0)}{d z_H d^2 \vec{p}_H} \Bigg |_{z_H=1}  + \frac{d \Phi^{\{Hg\}{\rm no \; plus}}_{PP}(x_H, \vec{p}_H,\vec q;s_0)}{d x_H d^2 p_H} .
\label{Eq:TermPlus}
\end{equation}
The pieces $d\tilde{\Phi}$, $d\Phi^{\{Hg\}(1-x_H)}$, and $d\Phi^{\{Hg\} {\rm real} \; P_{gg}}$ are free from the divergence for $z_H\to 1$ and therefore in their expressions the limit $s_{\Lambda} \rightarrow \infty$ can be safely taken, which means that $\theta(s_{\Lambda} - s_{PR})$ can be set to one. Moreover $d\tilde{\Phi}$ is also IR-safe and can be written in the more explicit form 
\begin{equation*}
    \frac{d \tilde{\Phi}^{\{Hg\}}_{PP}(x_H, \vec{p}_H,\vec q;s_0)}{d x_H d^2 p_H}
    =\int_{x_H}^1 \frac{d z_H}{z_H} f_g \left( \frac{x_H}{z_H} \right) \left[ \frac{d \Phi_{gg}^{ \{ H g \}}(z_H, \vec{p}_H, \vec{q};s_0)}{d z_H d^2 \vec{p}_H} - \frac{g^2  N}{4 (1-\epsilon) \sqrt{N^2-1}} \right.
\end{equation*}
\begin{equation}
    \left. \times \frac{\left| F_T ( 0, -\vec{p}_H^{\; 2}, m_{H}^2  ) \right|^2}{{ (2 \pi)^{D-1} (\vec{q}-\vec{p}_H)^2}} \left( \frac{z_H}{1-z_H} \vec{q}^{\; 2} + z_H (1-z_H) \vec{q}^{\; 2} + 2 (1-\epsilon) \frac{1-z_H}{z_H} \frac{(\vec{q} \cdot (\vec{q}- \vec{p}_H))^2}{(\vec{q} - \vec{p}_H)^2} \right) \right] .
\end{equation}
We finally show the explicit cancellation of rapidity divergences. The last term in eq.~(\ref{subtractions}) can be easily calculated from eq.~(\ref{Eq:HighRapidityLimitIF}) and gives 
\begin{equation*}
    \int_{x_H}^1 d z_H f_g (x_H)  \frac{d \Phi_{gg}^{ \{H g \}} (z_H, \vec{p}_H, \vec{q} ; s_0)}{d z_H d^2 \vec{p}_H} \bigg|_{z_H \rightarrow 1}
\end{equation*}
\begin{equation*}
=    \frac{g^2 |F_T ( 0, -\vec{p}_H^{\; 2}, m_{H}^2  )|^2 N}{4 (1-\epsilon) \sqrt{N^2-1} (2 \pi)^{D-1} } \frac{\vec{q}^{\; 2}}{(\vec{q}- \vec{p}_H)^2} \int_{x_H}^1 d z_H \frac{1}{(1-z_H)} f_g (x_H) \theta \left( s_{\Lambda} - \frac{(\vec{q}-\vec{p}_H)^2}{(1-z_H)} \right) 
\end{equation*}
\begin{equation}
    = \frac{g^2 |F_T ( 0, -\vec{p}_H^{\; 2}, m_{H}^2  )|^2 N}{4 (1-\epsilon) \sqrt{N^2-1} (2 \pi)^{D-1} } \frac{\vec{q}^{\; 2}}{(\vec{q}- \vec{p}_H)^2} f_g (x_H) \left[ \ln (1-x_H) - \frac{1}{2} \ln \left( \frac{\left[(\vec{q}-\vec{p}_H)^2 \right]^2}{s_{\Lambda}^2} \right) \right] \; .
\label{Phi_g_rap}
\end{equation}
Let us consider now the BFKL counter-term, given by the last term of eq.~(\ref{ImpactUnpro}),
\begin{equation}
   \frac{d \Phi^{\rm{BFKL \ c.t.}}_{gg}(z_H, \vec{p}_H, \vec q;s_0)}{d z_H d^2 p_H} = - \frac{1}{2} \int d^{D-2} q' \frac{ \vec{q}^{\; 2}}{\vec{q}^{\; '2}}  \frac{d \Phi_{gg}^{ \{H \} (0)} (\vec{q} \; ' )}{d z_H d^2 p_H}  \mathcal{K}^{(0)}_r (\vec{q} \; ', \vec{q} \; ) \ln \left( \frac{s_{\Lambda}^2}{(\vec{q} \; ' - \vec{q} \; )^2 s_0} \right) \; .
\end{equation}
Using eq.~(\ref{Eq:LoImpactD4-2EPart}) and eq.~(\ref{BornKer}), we find
\begin{equation}
    \frac{d \Phi^{\rm{BFKL \ c.t.}}_{gg}(z_H, \vec{p}_H, \vec q;s_0)}{d z_H d^2 p_H} \!=\! \frac{-g^2 |F_T ( 0, -\vec{p}_H^{\; 2}, m_{H}^2  )|^2 N}{8 (2 \pi)^{D-1} (1-\epsilon) \sqrt{N^2-1}} \frac{\vec{q}^{\; 2}}{(\vec{q}-\vec{p}_H)^2} \ln \left( \frac{s_{\Lambda}^2}{(\vec{q} - \vec{p}_H )^2 s_0}  \right) \!\delta (1-z_H)
\end{equation}
and, after convolution with the gluon PDF, we get
\[
 \frac{d \Phi^{{\rm{BFKL\ c.t.}}}_{PP}(x_H,\vec p_H,\vec q;s_0)}{d x_H d^2 p_H} = \int_{x_H}^1 \frac{d z_H}{z_H} f_g \left( \frac{x_H}{z_H} \right) \frac{d \Phi_{gg}^{{\rm{BFKL\ c.t.}}} (z_H, \vec{p}_H, \vec{q})}{d z_H d^2 \vec{p}_H} 
 \]
 \begin{equation}
= - \frac{g^2 |F_T ( 0, -\vec{p}_H^{\; 2}, m_{H}^2  )|^2 N}{8 (2 \pi)^{D-1} \sqrt{N^2-1}} \frac{\vec{q}^{\; 2}}{(\vec{q}-\vec{p}_H)^2} \frac{f_g (x_H)}{(1-\epsilon)} \ln \left( \frac{s_{\Lambda}^2}{(\vec{q} - \vec{p}_H )^2 s_0}  \right) \; .
 \label{BFKLct}
\end{equation}
When we combine the last term of eq.~(\ref{subtractions}), given in~(\ref{Phi_g_rap}), with the BFKL counter-term, given in~(\ref{BFKLct}), we obtain  
\begin{equation}
 \frac{d \Phi^{{\rm{BFKL}}}_{PP}(x_H, \vec{p}_H, \vec q;s_0)}{d x_H d^2 p_H} \equiv \frac{g^2 |F_T ( 0, -\vec{p}_H^{\; 2}, m_{H}^2  )|^2 N}{4 (2 \pi)^{D-1} (1-\epsilon) \sqrt{N^2-1}} \frac{\vec{q}^{\; 2}}{(\vec{q}-\vec{p}_H)^2} f_g (x_H) \ln \left( \frac{(1-x_H) \sqrt{s_0}}{|\vec{q} - \vec{p}_H|}  \right) \,.
 \label{CountFin}
\end{equation}
Note that this term is finite as far as the high-energy divergence is concerned. 
The BFKL term in eq.~(\ref{CountFin}) contains a soft singularity, which is expected to be cancelled when this contribution is combined with the virtual corrections, and a collinear singularity proportional to $\ln (1-x_H)$ which cancels an analogous one in eq.~(\ref{Eq:Term(1-zH)}). The remaining collinear singularity, associated with the initial state radiation, cancels when the proper NLO definition of the gluon PDF is employed. In particular, the term in eq.~(\ref{Eq:TermPlus}) produces the real part of the $P_{gg} (z_H)$ DGLAP splitting function~\cite{Celiberto:2022fgx}.   

\section{Conclusions and outlook}
\label{sec:conclusions}

In this paper we have calculated the next-to-leading order corrections to the impact factor for the Higgs boson production from a proton, due to  the emission of an extra parton in the proton fragmentation region, the so-called {\it real} corrections. We have used a finite value for the top quark mass, going therefore beyond the infinite-top-mass approximation which was adopted in previous calculations of the same object. This is the first step towards the calculation of the full Higgs impact factor, which will include also corrections coming from {\it virtual} radiation. When available, the Higgs impact factor will be used to build, with QCD in the high-energy limit with next-to-leading logarithmic approximation, new predictions for processes such as the inclusive forward Higgs production and the inclusive production of a forward Higgs and a backward identified object (a jet or a hadron) at the LHC and future hadron colliders, thus contributing to the exploration of the Higgs sector of the Standard Model in kinematic sectors going beyond the reach of a pure fixed-order approach, not supplemented by all-order resummations. \\

\noindent We checked explicitly that our result is compatible with gauge invariance and is exempt from rapidity divergences in the case of emission of a forward gluon. These are non-trivial tests of correctness of the result.
As for infrared divergences, their explicit cancellation can be shown only after the combination of real and virtual corrections and, {\it e.g.}, performing the projection onto the eigenfunctions of the leading BFKL kernel, {\it i.e.} after transferring the Higgs impact factor into the so-called $(n,\nu)$-representation (see Ref.~\cite{Celiberto:2022fgx} for details). We gave however some arguments supporting the fact that the result presented in this work has the correct infrared structure.

\acknowledgments

We thank Maxim A. Nefedov, Samuel Wallon, Renaud Boussarie, Lech Szymanowski, and Fulvio Piccinini for discussions. We are grateful to Vittorio Del Duca, Carlo Oleari, Carl R. Schmidt and Dieter Zeppenfeld for communications regarding the results of Ref.~\cite{DelDuca:2001fn}. The work of F.G.C. is supported by the Atracci\'on de Talento Grant n. 2022-T1/TIC-24176 of the Comunidad Aut\'onoma de Madrid, Spain. The work of L.D.R., G.G. and A.P. is supported by the INFN/QFT@COLLIDERS Project, Italy. The work of M.F. is supported by the Agence Nationale de la Recherche under the contract
ANR-17-CE31-0019. M.F. acknowledges support from the Italian Foundation “Angelo Della Riccia”. All pictures in this work have been drawn using JaxoDraw~\cite{Binosi:2008ig}. \\

\bibliographystyle{apsrev}
\bibliography{references}

\end{document}